\documentclass[]{aa} 
\usepackage{graphicx}
\usepackage{txfonts}

\usepackage{natbib}
\usepackage{amstext}

\usepackage{xspace}

\newcommand{\earth}{\hbox{$\oplus$}}
\newcommand{\mearth}{\ensuremath{M_\mathrm{\earth}}\xspace}
\newcommand{\tcomp}{\ensuremath{t_\mathrm{comp}}\xspace}

\newlength{\imwidth}
\begin{document}


\title{Giant planet formation: episodic impacts vs. gradual core growth}


\author{Christopher H. Broeg\inst{1}
\and
Willy Benz\inst{1}
} 
\institute{Center for Space and Habitability \& Physikalisches Institut, University of Bern,
 Sidlerstrasse 5, CH-3012 Bern, Switzerland\\
\email{broeg@space.unibe.ch}\thanks{please send all correspondence to Ch. Broeg}}

\date{Received ; accepted 17 November 2011} 

\abstract
{We describe the growth of gas giant planets in the
core accretion scenario.}
{ The core growth is not modeled as a gradual
accretion of planetesimals but as episodic impacts of large mass
ratios, i.e.\ we study impacts of 0.02 - 1 \mearth onto cores of 1-15
\mearth. Such impacts could deliver the majority of solid matter in 
the giant impact regime. We focus on the thermal response of the
envelope to the energy delivery. Previous studies have shown that
sudden shut off of core accretion can dramatically speed up gas
accretion. We therefore expect that giant impacts followed by periods
of very low core accretion will result in a net increase in gas
accretion rate. This study aims at modelling such a sequence of events
and to understand the reaction of the envelope to giant impacts in more detail. }
{To model this scenario, we spread the impact energy deposition over a time that is
long compared to the sound crossing time, but very short compared to
the Kelvin-Helmholtz time. The simulations are done in spherical
symmetry and assume quasi-hydrostatic equilibrium.}
{Results confirm what could be inferred from previous studies: gas can be accreted faster onto
the core for the same net core growth speed while at the same time rapid
gas accretion can occur for smaller
cores -- significantly smaller than the usual critical core
mass. Furthermore our simulations show, that significant mass
fractions of the envelope can be ejected by such an impact.
}
{
Large impacts are an efficient process to remove the accretion energy by
envelope ejection. In the time between impacts, very fast gas
accretion can take place.
 This process could
significantly shorten the formation time of gas giant planets. As an
important side-effect, the episodic ejection of the envelope will
reset the envelope composition to nebula conditions.
}


\titlerunning{GPF: episodic impacts vs. gradual core growth}
\authorrunning{Ch. Broeg \and W. Benz}
\maketitle


\graphicspath{{fig/}{./}}


\section{Introduction}
\label{sec:introduction}
We study the formation of gas giant planets in the core accretion scenario \citep{mizuno1980,1996Icar..124...62P,1986Icar...67..391B}. 
In this scenario, a planetary embryo grows by accreting from a swarm of planetesimals. At some point the embryo becomes massive enough to gravitationally attract a gaseous envelope. The growth process, both in terms of solid and gas accretion, is controlled by the planetesimal accretion, which is typically modeled as a gradual accretion of small planetesimals. However, in the giant collision phase under certain conditions, the accretion process could be dominated by relatively large impacts  \citep{1969Icar...10..109S}. This is confirmed by Monte-Carlo planet formation
models (T. Schr\"oter et al., in preparation) and recent results from
N-body simulations \citep{Raymond:2005p21625,Nimmo:2006p21624}. In such cases, while the collisions are less frequent, each one increases the mass of the protoplanet by a significant amount (typically of order 10\%).  This possibility led us to investigate the importance of the nature of the solid accretion process in the overall growth of giant planets. In particular, we want to investigate if episodic but large impacts result in changes  in the mass and structure of the envelope when compared to gradual core growth? 

A body of work investigates the importance of impacts and/or
core luminosity on the evolution of the envelope of giant planets. One
such study concerned itself with the possibility of stripping the
envelope of Uranus by an impact induced shockwave
\citep{1990Icar...84..528K}. However, the authors did not follow the
long-term evolution of the post impact planet and did not consider the
possibility of subsequent re-accretion of gas. \citet{2006ApJ...650.1150I} study the collision of two giant planets to explain the low envelope mass of HD~149026b.  Another study
\citep{2007A&A...466..717A} tried to assess  the effect of a large
impact on the long-term luminosity evolution of a giant planet with an
eye on its potential detection. Other studies investigated the
influence of the thermal energy content of the solid core as well as
the energy provided by its contraction on the overall evolution of the
luminosity \citep{2008A&A...482..315B}. Further studies  \citep{2007MNRAS.381..640P} concern the dynamic response of proto-planetary envelopes to a perturbation within an ideal gas approach. Recently, \citet{2010ApJ...720.1161L} have studied the merger of planetary embryos focussing on the re-distribution of heavy elements following the merger.

Most closely related to the problem at hand is the sudden core luminosity shut-off scenario: the evolution of the planet when the core luminosity is suddenly shut off. This has been studied in detail by \citet{2000ApJ...537.1013I} and \citet{Hubickyj2005415}. It is thus expected, that the periods of low core accretion in-between impacts will lead to massive gas accretion.

The effect of sporadic, relatively massive impacts during the growth phase of the core on the gas accretion has, however, not been studied in detail. Here we attempt to determine the thermal response of a gaseous envelope upon a sudden energy input delivered by a large impact to the core, and how such episodic events could modify the build-up of the envelope when compared to the nominal case of gradual accretion.

\section{Material and Methods}
\label{sec:material-methods}

\subsection{Structure equations}
\label{sec:structure-equations}

In the core accretion scenario in which we place our studies, a
growing giant planet is composed of a solid core surrounded by a
gaseous atmosphere. To model such a structure, we solve the standard
so-called equations of stellar structure. Please refer to table~\ref{tab:symbols} for the explanation of
all symbols. The equations are the same as in \citet{Broeg200915}, except for the envelope which is considered to be in quasi-hydrostatic equilibrium and the fact that we take into account its contraction: 
  \begin{eqnarray}
    \label{eq:1}
 \nabla \cdot F &=& \rho (\epsilon_{ac}- \dot q),\quad 
\text{or}
 \nonumber\\
\frac{\partial l}{\partial r} &=&  4\pi r^2 \rho (\epsilon_{ac}- \dot q),\quad \text{where}\\
 \dot q &=& T \dot s = c_p \dot T -\frac{\delta}{\rho} \dot P,\qquad
  l = 4\pi r^2 F. \nonumber 
  \end{eqnarray}

We assume the envelope to be of solar composition and use the equation of state from \citet{1995ApJS...99..713Sv}. Following the recommended
procedure for this equation of state, the effects of high-Z elements are accounted for by a somewhat enhanced He-mass fraction ($Y=0.3$).

For the opacity we use tabulated values: \citet{1985Icar...64..471P} (for $\lg T <2.3$) combined with molecular opacities from \citet{1994ApJ...437..879A} and high temperature opacities from \citet{1990ADNDT..45..209W}.

\subsection{Impact treatment}
\label{sec:impact-treatment}
An impact takes place on a timescale very short compared to the evolutionary timescale of the core and envelope; compared to this timescale it is essentially instantaneous. In addition, it is not spherically symmetric -- a three-dimensional analysis would be required to accurately model it and the investigation of various impact parameters would be required. Such studies, including following the evolution of the post-impact planet over several Kelvin-Helmholtz time scales are currently computationally prohibitive. We therefore opted for the study of a reduced and simplified problem: The thermal response of a spherically symmetric envelope in quasi-static equilibrium to an energy deposition onto the core corresponding to the energy delivered by the impact. 

Of course, this implicitly assumes that the impactor reaches the core
which is actually a good assumption for large impactors
\citep[see][]{2010ApJ...720.1161L}. The quasi-hydrostatic assumption,
on the other hand, is not correct if the impact is truly
instantaneous. However, provided that the envelope is not ejected by the
shockwave itself, the thermal energy release of the impact energy by
the core will not be instantaneous.  As shown
by \citet{1990Icar...84..528K}, in many cases the shockwave does not
unbind a significant fraction of the atmosphere. For these situations,
the major part of the impact energy is deposited directly into the core. We can very crudely estimate how fast this energy can
be released by the core as follows. Assuming the projectile is spread
entirely over the core, it will form a hot layer of constant
thickness. For a 10 \mearth core and a 0.02 \mearth impactor and
assuming a density of $5500\,\mathrm{kg/m^3}$ this layer will have a
thickness of 10 km. A layer of depth $D$ cools over a timescale given
by \(\tau_{cool} = D^2/a\) where \(a=\lambda / c_p \rho\) and
$\lambda$ is the thermal conductivity of the material. Using typical
values for the Earth mantle: $\lambda=50\,\mathrm{W m^{-1}  K^{-1}}$
\citep{Tang09032010}, $\rho=5500\,\mathrm{kg/m^3}$, $c_p=625\,
\mathrm{J kg^{-1}  K^{-1}}$ leads to a diffusion constant of $a\approx
1.5\cdot 10^{-5}\, \mathrm{m^2/s}$ and  a corresponding cooling time  of the order of  $\tau_{cool}=2\cdot10^5$ years.

 To derive this crude estimate,
we have assumed that energy was only transported by conduction and we
used values for the Earth mantle. This is certainly a lower limit for
the energy transport in the aftermath of the impact since other forms
of energy transport are possible, especially when the core becomes
partially molten and convection becomes important.

 In summary we conclude that in order to simulate 'impacts' the impact
 timescale must be no larger than this upper limit for the cooling 
 time.  To be on the safe side, we chose a ten times smaller
 value\footnote{or even smaller where possible} and
 set the impact timescale to $\tau_{imp}=10^4$
 years. 

 Based on the considerations above, we simulate the energy deposition due to the impact of a large planetesimal impact by a core accretion rate
$d M_z / d t$ modeled as a Gauss curve. We use the equivalent width  $\tau_{EW}$ of the Gauss curve as the timescale of the impact\footnote{The equivalent width of the Gauss curve is defined as the width of a rectangle having the full height of the Gaussian, that has the same surface area as the Gaussian. So $\tau_{EW}= \sigma \sqrt{2 \pi}$ where $\sigma$ is the standard deviation of the Gauss curve.}.
After the impact, the accretion rate is set to a low background value
of $\rm 10^{-10}\, \mearth / yr$. As a typical timescale we choose
$10^4$ years. This is 
significantly shorter than the Kelvin-Helmholtz time scale for
contraction\footnote{of the pre-impact configuration} but much larger than a
dynamical timescale. Sound travel time is of the order of one year for
extended protoplanets. In this way, we expect mach numbers below
$1/1000$ so that a quasi-hydrostatic treatment is justified. At the
same time, the impact timescale is much shorter than the
Kelvin-Helmholtz time scale and the planetary envelope has to adjust
its structure
to the energy input much faster than it can radiate the energy
away. Therefore the exact duration of the impact does not strongly
affect the results. These assumptions are confirmed by the actual computations.

With the scheme described above, we can compute the thermal response of the envelope to the energy deposited into the core. Again, we do not consider the effect of the initial shock wave generate by the impact. The validity of this assumption will be further discussed in section~\ref{sec:impact-treatment-1}.

\subsection{Implementation}
\label{sec:implementation}

To model this scenario, we have developed a new numerical code that solves the standard equations of stellar structure \citep{kippenhahn} on a self-adaptive 1-dimensional grid \citep{33102} using an implicit BDF for the time evolution. To simplify modeling mass accretion, we chose the radius $r$ as independent variable rather than mass. To handle the advection terms we use monotonized slopes after \citet{VanLeer1977276}. The equations are discretized using a finite-volume method on a staggered mesh and the resulting non-linear equation system is solved iteratively \citep{1964ApJ...139..306H}.

\begin{figure*}[tbp]
  \centering
  \includegraphics[width=17cm]{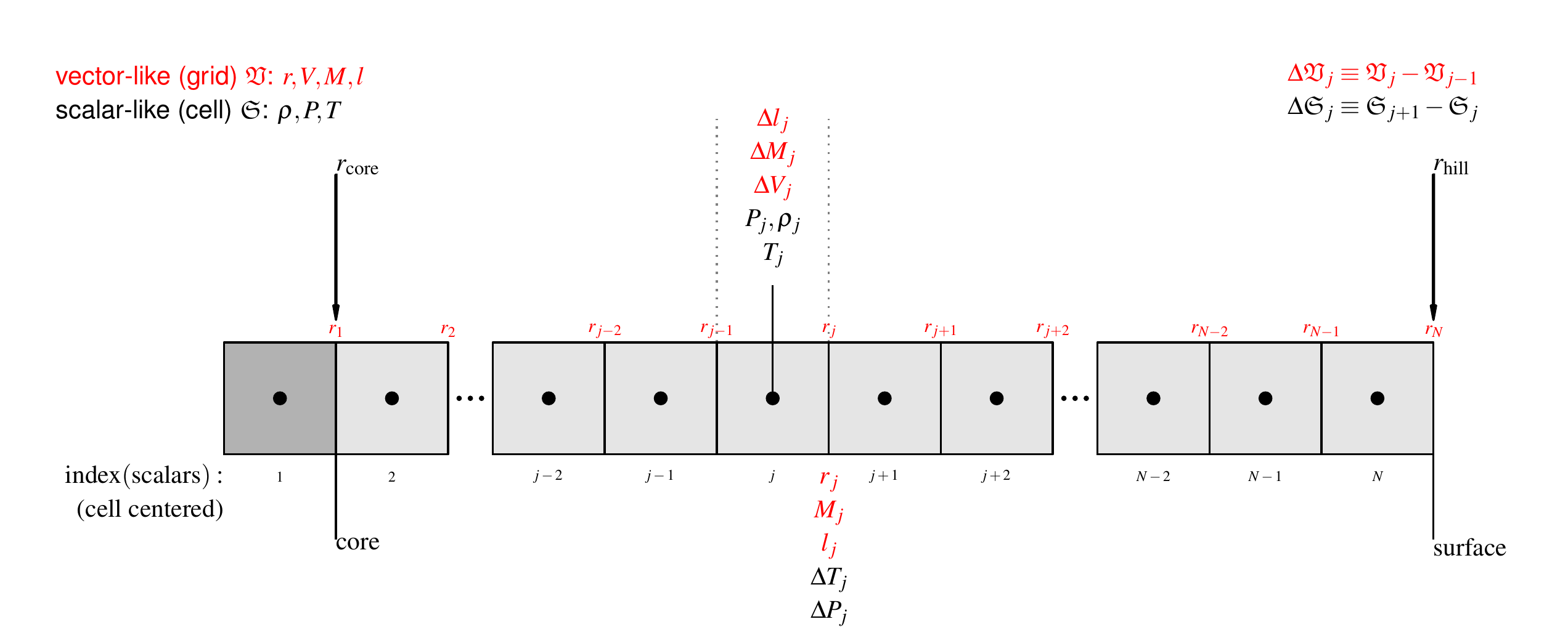}
  \caption{Layout of staggered mesh}
  \label{fig:grid}
\end{figure*}

In the following we present the discretized equations that we use. For brevity, we omit time-centering and the advection procedure. On the staggered mesh, scalar quantities ($P, T, \rho, \Delta V$) are cell centered, and vector like quantities ($r, V, M, u, l$) grid centered. This implies different delta operators for vector and scalar quantities, see Fig.~\ref{fig:grid} for the grid layout. Averaged quantities are indicated using intermediate indices: $P_{j\pm 1/2}=\frac{1}{2}(P_j+P_{j\pm 1})$.

The discretized equations are:

\subsubsection{conservation of mass}
\label{sec:conservation-mass}

\begin{equation}
  \Delta M_j = \rho_j \Delta V_j \nonumber
\end{equation}

\subsubsection{temperature gradient}
\label{sec:temperature-gradient}
\begin{equation}\label{eq:temperature-gradient}
  \Delta \ln T_j = \nabla_{j+1/2} \Delta \ln P_j
\end{equation}
with $\nabla$ calculated as:
\begin{equation}\label{eq:nabeff}
  \nabla \equiv \frac{d\ln T}{d\ln P} = {\rm min}(\nabla_\mathrm{rad}, \nabla_\mathrm{s}),
\end{equation}
i.e.\ the adiabatic temperature gradient $\nabla_\mathrm{s}$ or the temperature gradient as caused by radiative energy transport in the diffusion approximation, $\nabla_\mathrm{rad}$ -- whichever is smaller.  This corresponds to the use of zero entropy gradient convection and the application of the Schwarzschild-criterion. $\nabla_\mathrm{s}$ is directly given by the equation of  state, $\nabla_\mathrm{rad}$ is calculated as:
\begin{equation}
  \label{eq:radiation_diffusion}
 \nabla_\mathrm{rad} =  \frac{3}{64\pi \sigma G}\, \frac{\kappa l P}{T^4 M}
\end{equation}
\citep[see][]{mihalas}.

 In detail, $\nabla_{j+1/2}$ is calculated as:
\[
 \nabla_{j+1/2} = \nabla(P_{j+1/2}, T_{j+1/2}, M_j, l_j)\]
\[
\nabla(P, T, M, l) = \nabla\left[ \nabla_\mathrm{s}(P, T, M, l), \nabla_\mathrm{rad}(P, T, M, l)\right] 
\]
\[
\nabla(\nabla_\mathrm{s}, \nabla_\mathrm{rad}) =  \nabla_\mathrm{rad}
(1 - \theta) +  \nabla_\mathrm{s} \theta
\]
\[
\theta \equiv \frac{1}{2} \left[ \tanh \left(\frac{ \nabla_\mathrm{rad}-
    \nabla_\mathrm{s}}{\epsilon}\right) + 1 \right]; \qquad \epsilon=10^{-4}
\]
And eq~(\ref{eq:temperature-gradient}) is discretized as:
\begin{eqnarray}
&&[\log T_{j+1} - \log T_j] \nonumber \\\nonumber
-&&\nabla(P_{j+1/2}, T_{j+1/2}, M_j, l_j)\, [\log
P_{j+1} - \log P_j] = 0
\end{eqnarray}

\subsubsection{hydrostatic equilibrium}
\label{sec:hydr-equil}

\begin{equation}
    4 \pi r_j^2 \Delta(P_j + P^{rad}_j   ) = -G (m_j+M_c) \rho_{j+1/2} \Delta V_{j+1/2} / r_j^2
\end{equation}
where
\[
 \Delta V_{j+1/2} = \frac{4 \pi}{3} (r^3_{j+1/2} - r^3_{j-1/2}),\quad
 \text{and}\quad
P^{rad} = \frac{a}{3} T^4.\nonumber
\]

\subsubsection{energy equation}
\label{sec:energy-equation}
In our advection scheme we make use of the connection between mass
flow and relative  velocity: \( \widetilde{\rho} u^{rel} 4 \pi r^2 =
-\frac{\delta m}{\delta t}.\) The resulting equation reads:
\begin{eqnarray} 
  \Delta \widehat {l_j} = -\widehat {c_{p,j}} \left\{\delta (\rho T \Delta V)_j/ \delta t - \Delta_j(\widetilde{T} \frac{\delta m}{\delta t})\right\}\nonumber\\
  + \frac{\widehat{\delta_j}}{\widehat{\rho_j}} \left\{\delta(\rho P \Delta V)_j/ \delta t - \Delta_j(\widetilde{P} \frac{\delta m}{\delta t})\right\}
\end{eqnarray}
where the tilde indicates quantities that are advected via van Leer monotonized slopes and the hat indicates time-centered quantities. For the results presented in this article we set the time-centering parameter $\theta=1$ (fully implicit).

\subsubsection{Grid equation}
\label{sec:grid-equation}
The use of the radius as independent variable facilitates calculation of mass flow through the outer boundary and allows the calculation of "detached" planets, cases for which the mass as variable becomes singular. However, within this formalism the existence of strong pressure gradients and gradients in the opacity require a self-adaptive grid which adapts to the evolving structure of the planet. We use a modified version of \citet{33102}. The adaptive grid is defined by specifying the point concentration $n$, defined as:
\begin{equation}
   n_j = \frac{\chi_j}{\Delta r_j}\quad \text{where}\quad\chi_j:\text{ typical scale.}
\end{equation}
Setting $n_j$ constant implies an equidistant grid. Using a fixed scale implies an equidistant grid in linear space, using a local scale implies log-equidistant spacing. One typically takes the local average radius: $\chi_j=r_j+r_{j-1}$. The factor 0.5 is dropped because it makes no difference for the relative spacing.

More general, we want to set the local point concentration proportional to some requested resolution $\mathcal{R}$:
\begin{equation}
n \propto \mathcal{R}
\end{equation}
One way to prescribe this proportionality in a discretized way is to set:
\begin{equation}
\frac{n_j}{\mathcal{R}_j} = \frac{n_{j+1}}{\mathcal{R}_{j+1}}
\end{equation}
How to prescribe the resolution is a free choice.  At the moment we use the length of the path in multidimensional space. Appropriate weights $g_j$ must be used because the different physical quantities have different physical units and can differ quite dramatically in dynamical range. We use the following required resolution:
\begin{equation}
\mathcal{R}_j^2 = 1 + n_j^2 \sum_{l=1}^M\left[g \left(\frac{\Delta f_j}{\overline{f_j}} \right)\right]_l
\end{equation}
where the index $l$ runs over the physical quantities that should be well-resolved in path-length (i.e.\ mass, temperature), $(f_j)_l$ represents each quantity at position $j$ on the grid, and $\bar{f}_j$ is the corresponding local scale (using a local average for $\overline{f_{j}}$ gives log-equidistant spacing along the path). The $g_l$ are the relative weights of the different physical quantities.

For the calculations presented here, we use \(m, P, T\) to specify the required resolution with the weights 1.0, 0.1, and 1.0, respectively. The mean values are calculated using the harmonic mean.

While the last two equations are, in principle, sufficient to define the grid, dramatic change in resolution can occur from one grid point to the next. Since numerically this leads to significant errors, we limit limit the maximum change in spacing from one point to its direct neighbors. This is called spacial smoothing. To achieve this, we follow Dorfi and apply a diffusive process on the point density: we replace $n_j$ by $\hat n_j$:

\begin{equation}
\hat n_j = n_j - \alpha_g(\alpha_g+1)(n_{j-1} - 2 n_j + n_{j+1})
\end{equation}
and require:
\begin{equation}
\frac{\hat n_j}{\mathcal{R}_j} = \frac{\hat n_{j+1}}{\mathcal{R}_{j+1}}
\end{equation}
$\alpha_g$ gives the strength of the smoothing. To allow a maximal change of 30\% from one cell to the next we set $\alpha_g=2$.

In addition to limiting sharp spatial change in resolution, sudden changes of the grid during time evolution need also to be avoided.  For this, we use a similar procedure called temporal smoothing. 

This is obtained by replacing the point density by the \emph{temporally smoothed} quantity $\widetilde{n}_j$:
\begin{equation}
\widetilde{n}_j = \hat n_j + \frac{\tau_{grid}}{\delta t} (\hat n_j - \hat n_j^{old})
\end{equation}
where $\tau_{grid}$ is the grid adaption time-scale and require instead:
\begin{equation}
\frac{\widetilde{n}_j}{\mathcal{R}_j} = \frac{\widetilde{n}_{j+1}}{\mathcal{R}_{j+1}}.
\end{equation}
For the computations we set $\tau_{grid}=100$ years.

\subsection{Boundary conditions}
\label{sec:boundary-conditions}

To specify the inner boundary condition for the core luminosity, we also integrate the gravitational potential. The equation reads:
\begin{equation}
  \label{eq:2}
  r^2_{j+1/2} \Delta \Phi_j = G (m_{j+1/2}+M_c) \Delta r_j.
\end{equation}

In total, we have 6 non-linear equations for the unknown quantities \(r, P, T,\Phi, m, l\). Due to the van Leer advection scheme, we end up with a stencil of
\(-3..2\), i.e.\ the Jacobi Matrix has a banded structure with 17 upper and 23 lower non-zero diagonals. Near the boundaries, we use donor cell advection with a stencil of $-1..1$.

For the 5 discretized differential equations we have to specify the initial values and one boundary condition per equation. For the grid equation we must specify both boundaries. In total, we need to specify 7 boundary conditions. They are:
\begin{eqnarray*}
  \label{eq:3}
  r_1 &=& r_{core}\\
 m_1 &=& 0\\
l_1&=& -\dot M_z \Phi_1\\
  r_N &=& r_{hill}\\ 
P_N &=& P_{neb}\\
  T_N &=& T_{neb}\\ 
\Phi_N &=& 0 \\
\end{eqnarray*}
where the index 1 stands for the inner boundary and $N$ for the outer. The outer radius is given by the hill radius $r_{hill}=a\sqrt[3]{{M}/({3 M_{\ast}})}$; $\dot M_z$ is the accretion rate of solids. For the initial values we start with a static model at a tiny core size.

\subsection{Code Verification}
\label{sec:code-verification}
\begin{figure}[thbp]
  \centering
  \includegraphics[width=\imwidth]{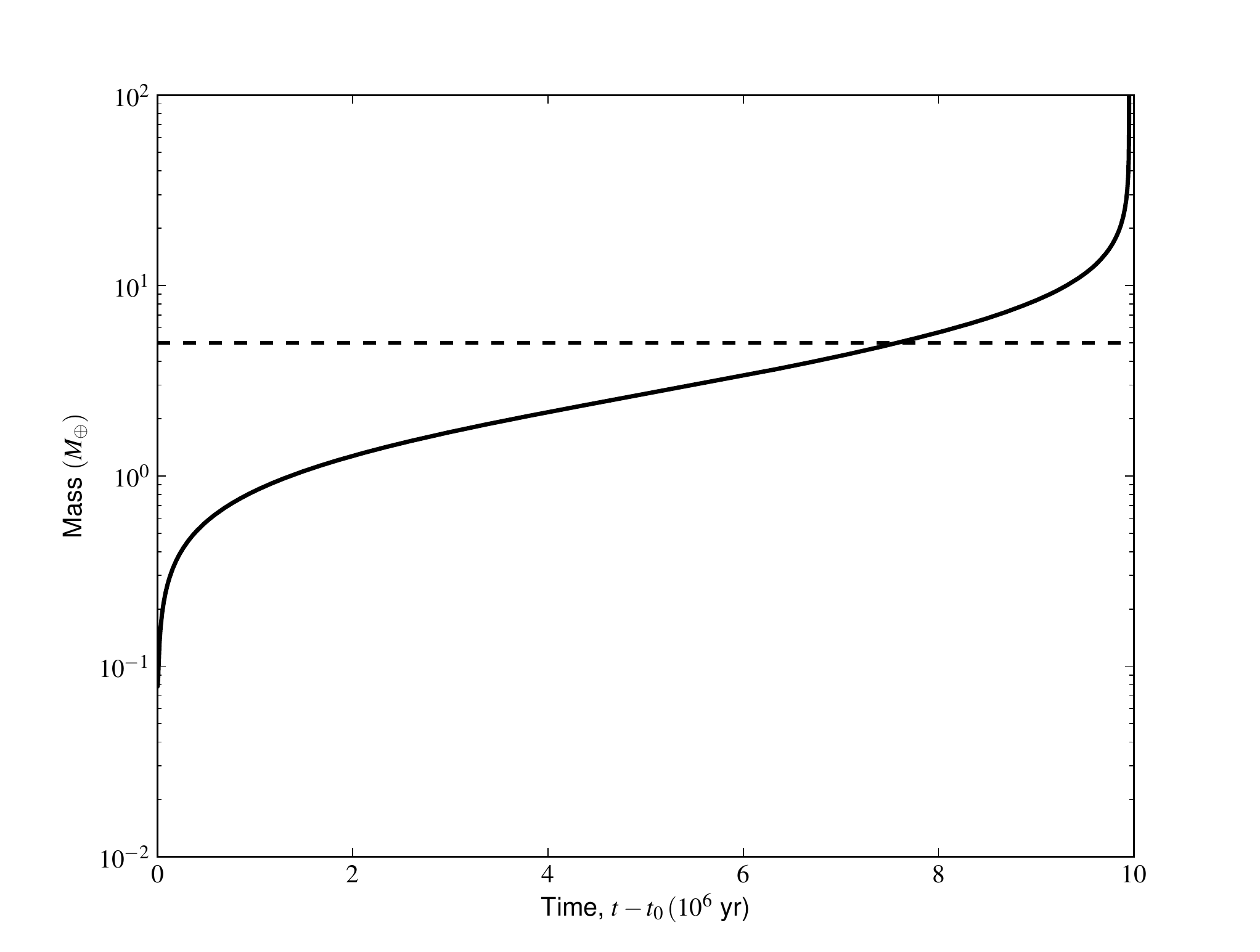}
  \includegraphics[width=\imwidth]{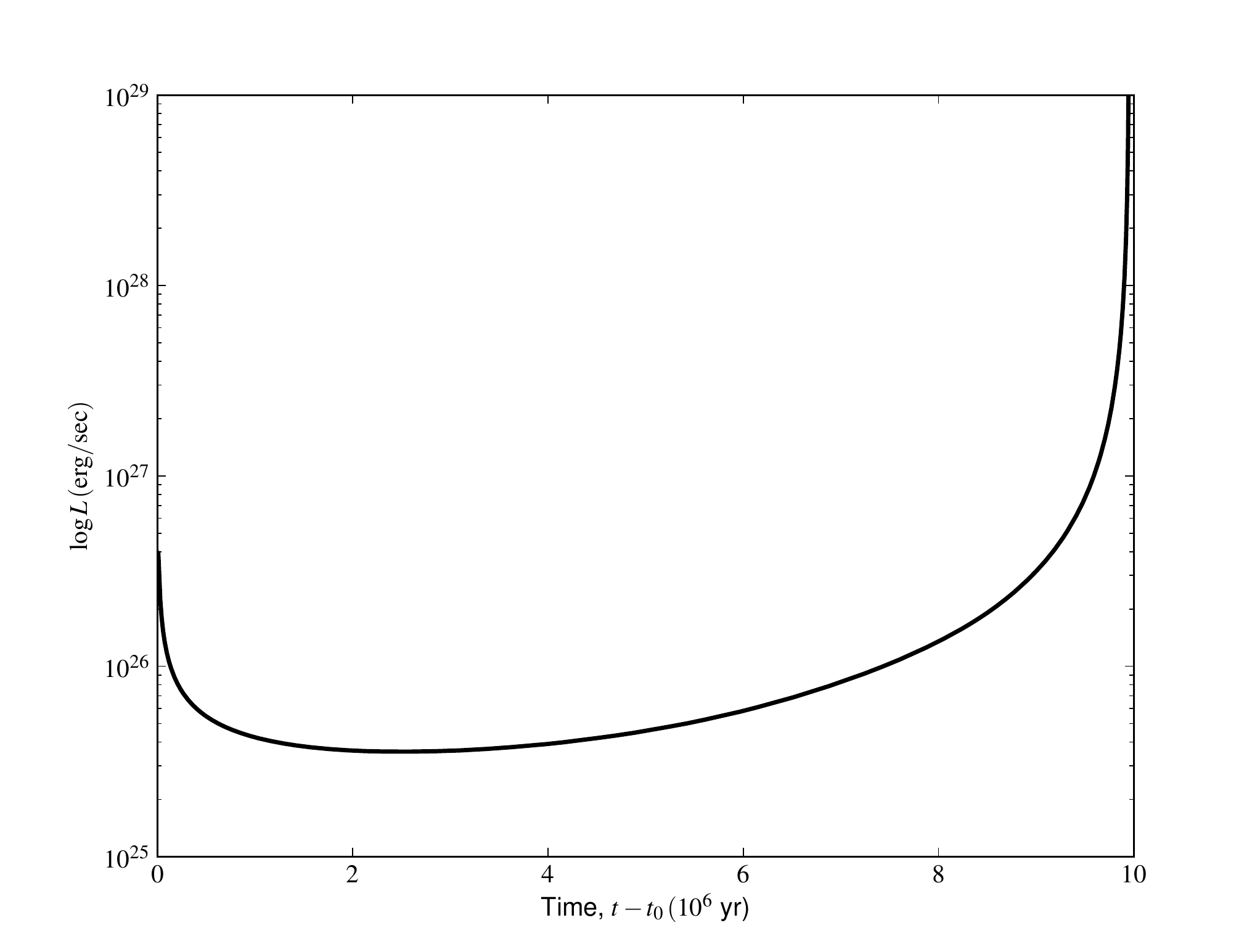}
  \caption{Time evolution of the envelope after core accretion is
    halted in the case of $M_{core}= 5 M_{\oplus}$. (a) shows the
    envelope mass $M_{env}$ (solid line) and the core mass $M_{core}$
    (dashed line) and (b) presents the luminosity $L$. This is in good
  agreement with  \citet{2000ApJ...537.1013I}, Fig. 2, even though the
used opacities from \citet{2005ApJ...623..585F} (g98.7.02.tron) are
not exactly equivalent.}
  \label{fig:ikoma}
\end{figure}

To verify the code we have compared the results extensively. For
static models ($l=const$) we have compared to our previous, well tested, shooting method code \citep{Broeg200915}. The results
were identical well below percent level. To test the luminosity
equation, we have compared to evolutionary tracks of HD209458b of
Tristan Guillot (priv. comm.) and found good agreement to 10 Gyrs.  
We also compared with CoRoT9b evolutionary tracks and we get a best
fit for 10 Earth masses of solids in the core, again in good agreement
with established calculations \citep{2010Natur.464..384D}. As a final
test we have repeated the calculations of \citet{2000ApJ...537.1013I}
where the core accretion is stopped for a 5 $M_\oplus$ core, see Fig.~\ref{fig:ikoma}.

\subsection{Procedure}
\label{sec:procedure}
\begin{figure}[thbp]
  \centering
  \includegraphics[width=\imwidth]{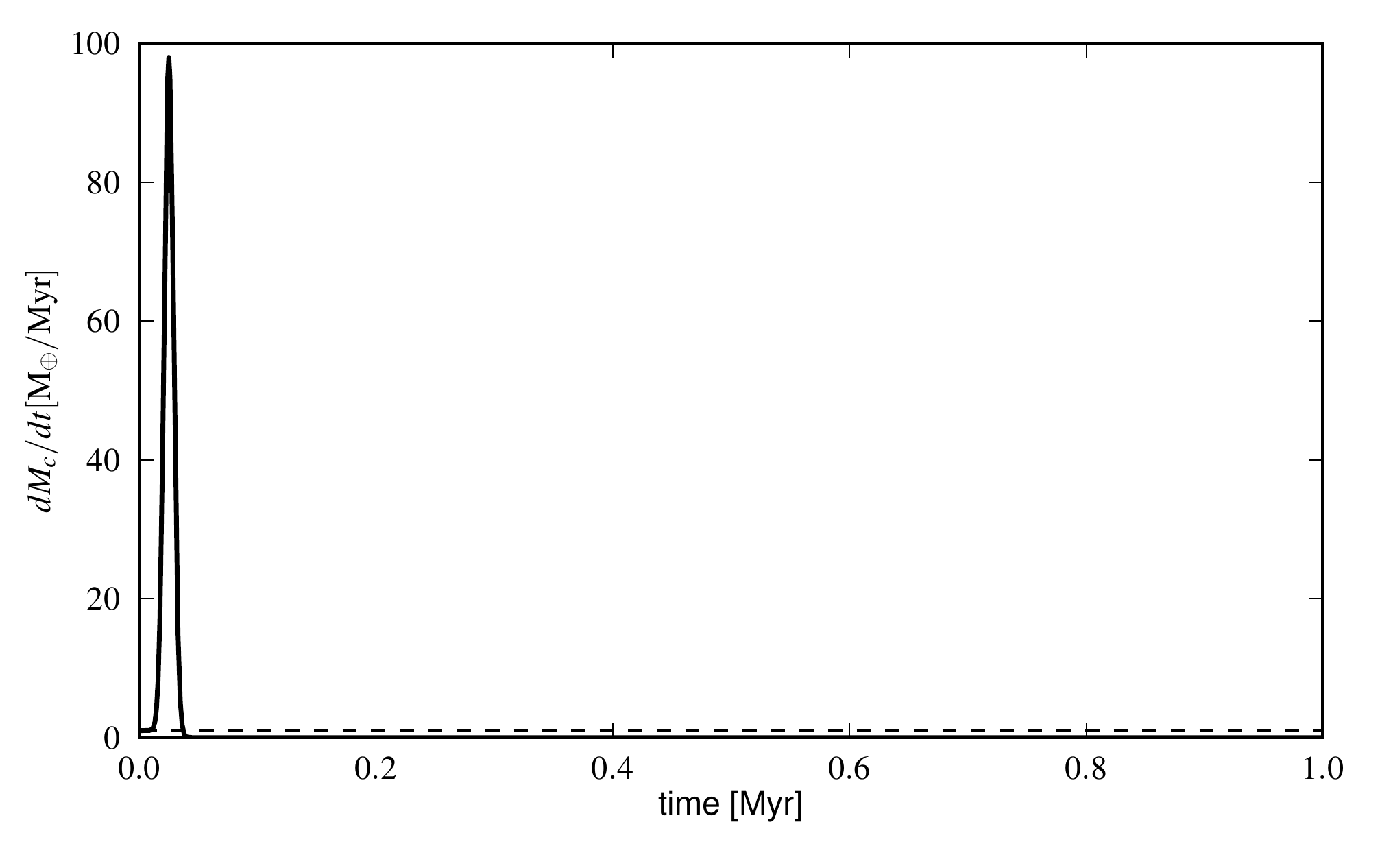} 
  \includegraphics[width=\imwidth]{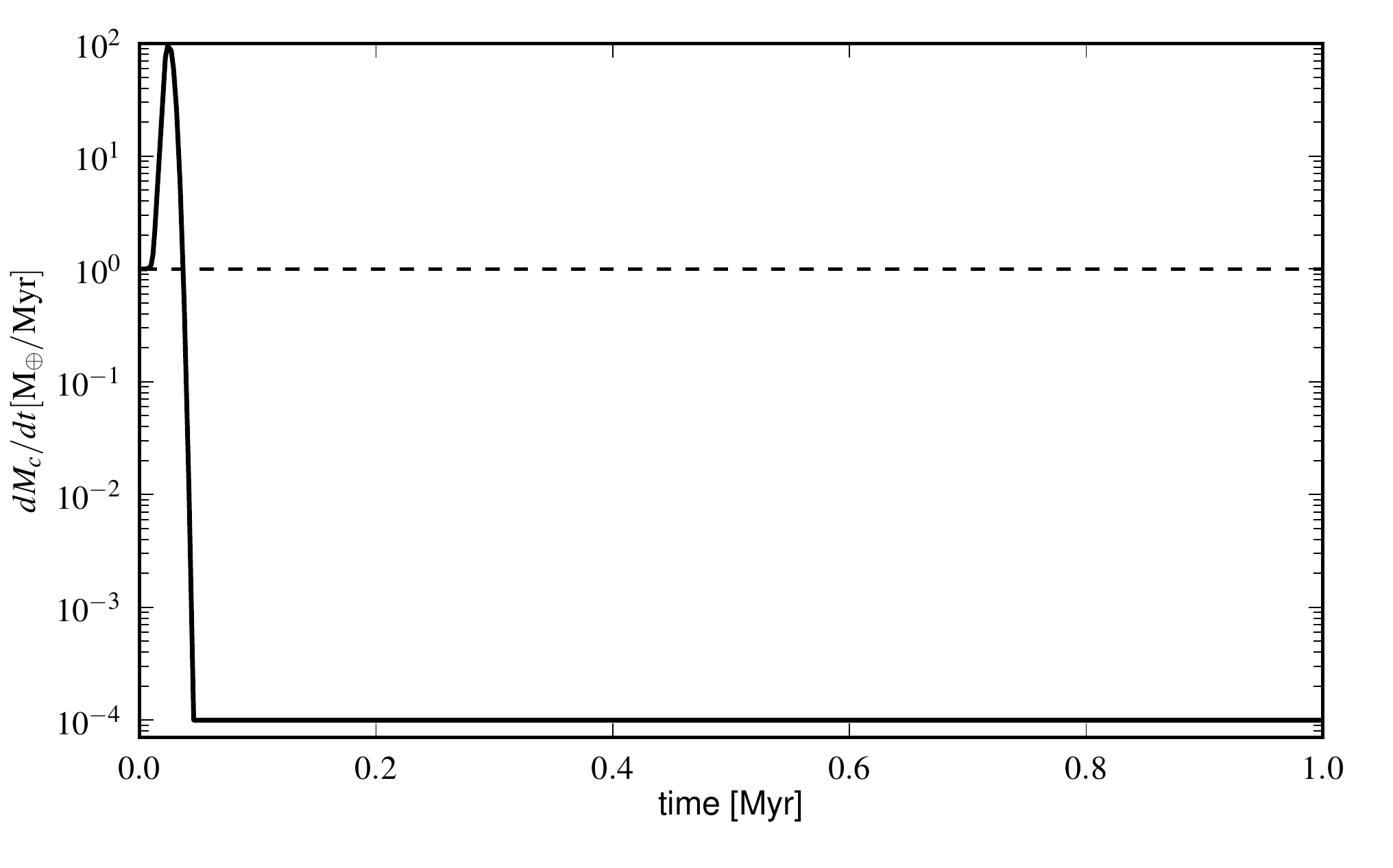}
  \caption{Simulation of one impact (here: 1 \mearth in $10^4$ yr): the core accretion rate is described by a Gaussian curve. The dashed curve shows the nominal case of constant accretion. For clarity both logarithmic and linear scale are shown.}
  \label{fig:impact}
\end{figure}

To study the effect of episodic impacts, we have chosen the following
procedure. A planetary embryo is placed into a static minimum mass solar nebula \citep{1981PThPS..70...35H} with a semi major axis of 3 AU in orbit around a sun-like star. The core grows gradually with an accretion rate of $10^{-6}\, \mearth/yr$ until the impact takes place. After the impact, a period of low accretion (background rate) is assumed until a reference core undergoing gradual accretion reaches the same mass. Fig.~\ref{fig:impact} shows an example of such a time sequence. 
Important parameters are the target mass, and the mass deposited by the impact. To investigate the various outcomes following changes in these parameters, we carried out a number of calculations for which we list the characteristics in Table~\ref{tab:impact}. In all cases, we assumed $10^4$ years for the characteristic impact energy deposition timescale
\begin{table}[tbp]
  \centering
  \caption{List of simulations performed with their corresponding parameters.}
  \label{tab:impact}
  \begin{tabular}{rl}
    \hline
parameter & value \\ 
\hline
target mass & 1, 2, 3, .. 15 \mearth\\
impact mass & 0.02, 0.1, 0.5, 1 \mearth\\
\hline
  \end{tabular}
\end{table}
except for the smallest impacts (0.01 \mearth) for which we have used $10^3$ years\footnote{Test computations show that the result is weakly dependent upon the assumed impact timescale. Example: For a 10 \mearth target and 0.5 \mearth impactor the final envelope mass changes from 2.78 to 2.90 adopting an impact timescale of $10^4$ or $10^3$ years, respectively. However, the mach number of the gas outflow reaches values of 0.05 in the latter case. In this regime, our
quasi-hydrostatic approximation starts to break down.}.

We end the computations when the reference core which accretes at the nominal rate of $10^{-6}\, \mearth/yr$ has reached the same mass. At this time, defined as the comparison time  \tcomp, we compare the episodic case (EC) with the nominal case (NC).  Key quantities of interest we use in this comparison are the envelope mass and the envelope accretion rate $d M_{env}/dt$. 
 
\section{Results}
\label{sec:results}
In this section, we  present the results from our series of
computations. We  begin with the presentation of the data and defer
the discussion to section~\ref{sec:discussion}. For each of the four
impactor masses, we summarize the result in a table
(Tables~\ref{tab:i002}--\ref{tab:i1}). In each table, there is one row
for each target mass listing the following quantities: the NC envelope
mass at \tcomp ($M_{\rm env}^0$), the EC envelope mass compared to the NC ($M_{\rm env}/M_{\rm env}^0$), the NC gas accretion rate at \tcomp ($dM_{\rm env}^0/dt $), the EC gas accretion rate ($\dot M_{\rm env} /\dot M_{\rm env}^0$) , the EC total luminosity compared to the NC ($L/L^0 $), the envelope mass fraction lost ($-\Delta M_{\rm env}[\%]$), and the ratio of the impact energy to the total binding energy of the envelope ($-E_{\rm imp}/E_{b}$). The ejected envelope fraction is calculated by comparing the envelope before the impact with the smallest occurring
envelope mass after the impact.

The impact energy $E_{\rm imp}$ is  the gravitational energy
liberated by moving the impactor from the hill sphere to the core of
the target:
\begin{equation}
  \label{eq:4}
  E_\mathrm{imp} = - \Phi(r_c) M_\mathrm{imp}.
\end{equation}
The total binding energy of the envelope is defined as the sum of
 gravitational  and  internal energy of the
envelope:
\begin{equation}
  \label{eq:6}
  E_b = E_\mathrm{grav} + E_\mathrm{int}\\
\end{equation}
which are calculated as:
\begin{eqnarray}
  \label{eq:5}
  E_\mathrm{grav} & =& - \sum_{j=2}^{N} G\frac{M_c + M_j}{r_j} \rho_j dV_j \\
  E_\mathrm{int} &=&  \sum_{j=2}^{N} e(T_j, P_j)\, \rho_j dV_j,
\end{eqnarray}
where $e$ is the internal energy as given by the equation of
state. Both impact energy and binding energy are determined at the
time of the impact for the yet unperturbed envelope.

\begin{figure}[tbp]
  \centering
\includegraphics[width=0.49\textwidth]{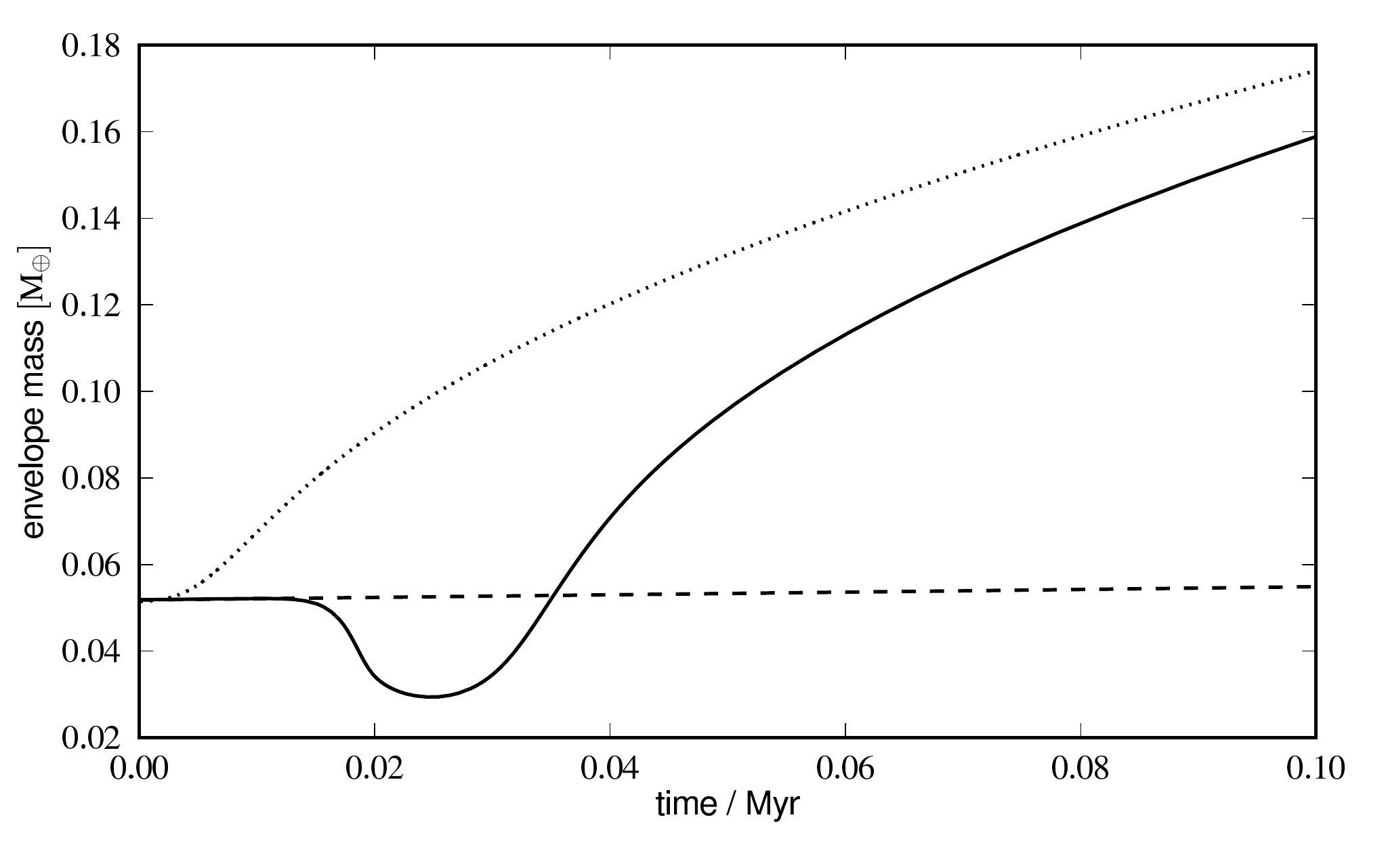} 
\includegraphics[width=0.49\textwidth]{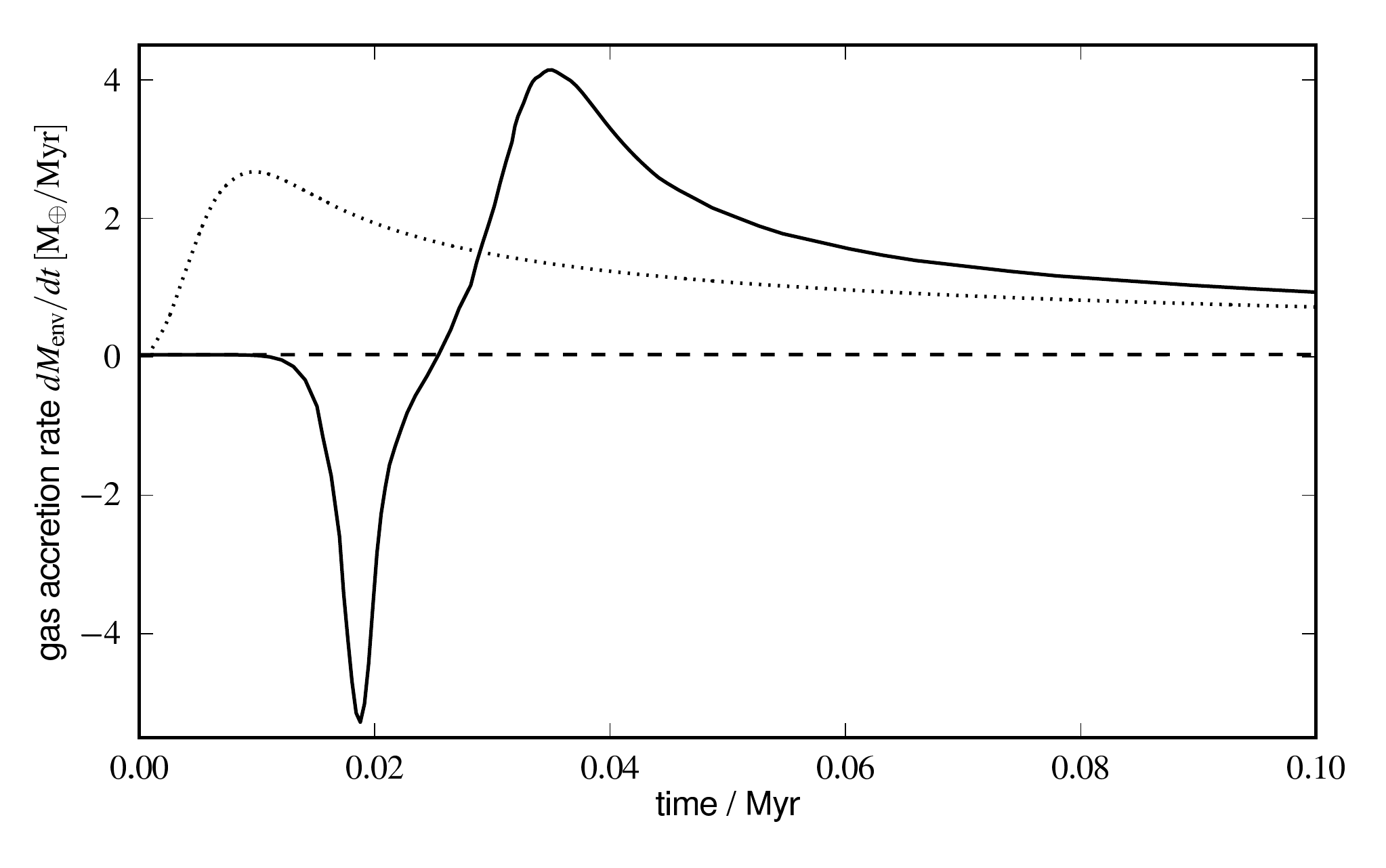} 
  \caption{Accretion of 1/10 \mearth by a core of 5 \mearth: envelope
    mass (top) and envelope accretion rate (bottom).
    Nominal case (constant $\dot M_z$, dashed line) vs. sudden accretion of
    1/10 \mearth (solid line) vs. shut-off of core luminosity (dotted
    line). The 
    timescale for impact and shut-off is 0.01 Myr. The offset of
    impact to shut-off comes from the fact that the core luminosity
    first rises and then decreases causing a delay. Also the
    mathematical prescription for core-luminosity differs slightly for
    shut-off and impact. In this case the 
    shut-off case is more massive at the end. This is not so for
    larger impacts because the larger core mass post impact leads to
    larger accretion rates and eventually larger envelope mass. Note
    that the peak accretion rate is larger after the impact than after
  the shut-off. }
  \label{fig:5Mc_env}
\end{figure}

Fig.~\ref{fig:5Mc_env} shows an example of the evolution of the
envelope mass and gas accretion rate as a function of time for a 5
\mearth core having suffered an impact with a 0.1 \mearth impactor
(solid line) and for the same core undergoing regular accretion
(dashed line). At first, the impact leads to a loss of envelope
mass. However, after a relatively short period in time, the accretion
of gas resumes at a rate much higher than in the nominal case. By the
time the nominal core reaches the same mass, the gas envelope accreted
by the EC is about three times more massive. In this case the shut-off
case (dotted line) accretes more gas because it reacts immediately to the shut-off
of core luminosity while the gas ejection takes a certain amount of
time. Afterwards, however the gas accretion rate in the EC is higher
because of the slightly larger core mass. 

\begin{table*}[hptb]
\newcommand\T{\rule{0pt}{2.6ex}}
  \centering
  \caption{End-of-calculation data for impacts of 0.02 \mearth for different core sizes
    compared to nominal case (NC).
    The NC accretes planetesimals at a constant rate of
    $10^{-6}\, \mearth/\rm{yr}$. Values are given at the end of
    computation when the core masses of nominal and science case are
    equal. All values are given with respect to the NC. NC values of
    envelope mass and envelope accretion rate are
    given in units of \mearth and $\mearth/\rm{yr}$,
    respectively. Target mass in \mearth. The last two columns give the ejected envelope fraction and the ratio of the impact energy to the binding energy of the envelope.}
\label{tab:i002}
  \begin{tabular}{cccccccc}
    \hline
$M_c^{\rm target}$  \T & $M_{\rm env}^{0}$  & $M_{\rm env}/M_{\rm env}^0$  & $dM_{\rm env}^0/dt $  & $\dot M_{\rm env} / \dot M_{\rm env}^0$ & $L/L^0 $ & $-\Delta M_{\rm env}[\%]$ & $-E_{\rm imp}/E_{b}$ \\\hline
1 & 0.00 & 1.22 & 4.49e-09  &  5.7& 0.0112&   1.3& -272.1 \\
2 & 0.01 & 1.58 & 5.73e-09 & 28.3 & 0.072 &   7.7&   840\\
3 & 0.02 & 1.91 & 9.26e-09 &   52 & 0.182 &  18.9&  31.3\\
4 & 0.03 & 1.97 & 1.69e-08 & 61.4 & 0.343 &  34.8&  7.96\\
5 & 0.05 & 1.80 & 3.00e-08 & 63.2 & 0.553 &  50.0&  2.99\\
6 & 0.09 & 1.54 & 5.05e-08 & 61.2 & 0.813 &  59.2&  1.39\\
7 & 0.16 & 1.30 & 7.97e-08 & 55.6 & 1.11 &  45.2&  0.74\\
8 & 0.26 & 1.12 & 1.24e-07 & 48.4 &  1.4 &  26.5& 0.433\\
9 & 0.41 & 1.01 & 1.81e-07 &   39 & 1.58 &  15.6& 0.272\\
10 & 0.63 & 0.97 & 2.64e-07 & 23.7 & 1.55 &   9.1& 0.179\\
11 & 0.96 & 0.96 & 3.76e-07 & 12.5 &  1.4 &   5.3& 0.122\\
12 & 1.43 & 0.97 & 5.60e-07 & 5.79 & 1.24 &   2.9& 0.0853\\
13 & 2.16 & 0.98 & 8.92e-07 & 2.41 & 1.13 &   1.3& 0.0602\\
14 & 3.39 & 0.98 & 1.65e-06 &    1 & 1.06 &   0.3& 0.0421\\
15 & 6.49 & 0.97 & 5.96e-06 & 0.866 & 0.999 &   0.0& 0.0266\\
\hline
  \end{tabular}
\end{table*}
\begin{table*}[hptb]
\newcommand\T{\rule{0pt}{2.6ex}}
  \centering
  \caption{End-of-calculation data for impacts of 0.1 \mearth. Same conventions as Table~\ref{tab:i002}.}
\label{tab:i01}
  \begin{tabular}{cccccccc}
    \hline
$M_c^{\rm target}$  \T & $M_{\rm env}^{0}$  & $M_{\rm env}/M_{\rm env}^0$  & $dM_{\rm env}^0/dt $  & $\dot M_{\rm env} / \dot M_{\rm env}^0$ & $L/L^0 $ & $-\Delta M_{\rm env}[\%]$ & $-E_{\rm imp}/E_{b}$ \\\hline
1 & 0.00 & 1.42 & 4.54e-09 & 3.25 & 0.00674 &   0.0& -1.36e+03\\
2 & 0.01 & 2.16 & 5.92e-09 & 14.8 & 0.0378 &   4.9& 4.2e+03\\
3 & 0.02 & 2.83 & 9.62e-09 &   26 & 0.0918 &  15.3&   157\\
4 & 0.03 & 3.07 & 1.75e-08 & 30.1 & 0.169 &  29.6&  39.8\\
5 & 0.05 & 2.90 & 3.17e-08 & 29.4 & 0.271 &  43.3&    15\\
6 & 0.10 & 2.55 & 5.29e-08 & 28.6 & 0.398 &  53.7&  6.95\\
7 & 0.17 & 2.19 & 8.45e-08 & 26.2 & 0.549 &  60.8&   3.7\\
8 & 0.27 & 1.86 & 1.28e-07 & 24.8 & 0.727 &  64.6&  2.17\\
9 & 0.43 & 1.59 & 1.89e-07 & 23.2 & 0.929 &  61.8&  1.36\\
10 & 0.66 & 1.37 & 2.68e-07 & 21.8 & 1.15 &  50.6& 0.893\\
11 & 0.99 & 1.20 & 3.88e-07 & 19.2 & 1.36 &  36.1&  0.61\\
12 & 1.48 & 1.08 & 5.77e-07 & 16.2 & 1.55 &  24.1& 0.426\\
13 & 2.23 & 0.99 & 9.23e-07 & 11.7 & 1.66 &  15.1& 0.301\\
14 & 3.53 & 0.94 & 1.75e-06 & 5.79 & 1.55 &   8.1&  0.21\\
15 & 7.06 & 0.91 & 7.39e-06 & 1.23 & 1.13 &   0.0& 0.133\\
\hline
  \end{tabular}
\end{table*}
\begin{table*}[hptb]
\newcommand\T{\rule{0pt}{2.6ex}}
  \centering
  \caption{End-of-calculation data for impacts of 0.5 \mearth. Same conventions as
    Table~\ref{tab:i002}. An asterisk indicates premature end of
    calculation due to rapid gas accretion.}
\label{tab:i05}
  \begin{tabular}{cccccccc}
    \hline
$M_c^{\rm target}$  \T & $M_{\rm env}^{0}$  & $M_{\rm env}/M_{\rm env}^0$  & $dM_{\rm env}^0/dt $  & $\dot M_{\rm env} / \dot M_{\rm env}^0$ & $L/L^0 $ & $-\Delta M_{\rm env}[\%]$ & $-E_{\rm imp}/E_{b}$ \\\hline
1 & 0.01 & 2.73 & 4.91e-09 & 2.33 & 0.00537 &   0.1& -6.79e+03\\
2 & 0.01 & 4.71 & 7.02e-09 &  6.7 & 0.0186 &   6.0& 2.1e+04\\
3 & 0.02 & 6.16 & 1.23e-08 & 9.38 & 0.0404 &  17.6&   783\\
4 & 0.04 & 6.56 & 2.24e-08 & 10.4 & 0.072 &  34.0&   199\\
5 & 0.07 & 6.20 & 3.93e-08 & 10.7 & 0.115 &  50.2&  74.8\\
6 & 0.12 & 5.57 & 6.33e-08 & 10.8 & 0.172 &  62.8&  34.8\\
7 & 0.20 & 4.93 & 9.99e-08 & 10.8 & 0.247 &  71.5&  18.5\\
8 & 0.33 & 4.37 & 1.48e-07 & 11.4 & 0.346 &  77.4&  10.8\\
9 & 0.51 & 3.92 & 2.20e-07 & 11.7 & 0.479 &  81.4&  6.79\\
10 & 0.77 & 3.55 & 3.13e-07 & 12.7 & 0.664 &  84.4&  4.46\\
11 & 1.16 & 3.27 & 4.54e-07 &   14 & 0.948 &  86.7&  3.05\\
12 & 1.74 & 3.07 & 6.94e-07 & 16.5 & 1.46 &  88.5&  2.13\\
13 & 2.65 & 3.01 & 1.16e-06 & 21.3 & 2.87 &  89.0&  1.51\\
14 & 4.40 & 5.03 & 2.56e-06 &  152 & 32.8 &  80.8&  1.05\\
15 & *12.28 & *11.79 & *4.64e-05  & *3.41e+06& * 104&*  52.6&*   0.7 \\
\hline
  \end{tabular}
\end{table*}
\begin{table*}[hptb]
\newcommand\T{\rule{0pt}{2.6ex}}
  \centering
  \caption{End-of-calculation data for impacts of 1 \mearth. Same conventions as
    Table~\ref{tab:i002}. An asterisk indicates premature end of
    calculation due to rapid gas accretion.}
\label{tab:i1}
  \begin{tabular}{cccccccc}
    \hline
$M_c^{\rm target}$  \T & $M_{\rm env}^{0}$  & $M_{\rm env}/M_{\rm env}^0$  & $dM_{\rm env}^0/dt $  & $\dot M_{\rm env} / \dot M_{\rm env}^0$ & $L/L^0 $ & $-\Delta M_{\rm env}[\%]$ & $-E_{\rm imp}/E_{b}$ \\\hline
1 & *0.01 & *4.84 & *5.71e-09  & *2.97& *0.00729&*   0.0&* -13602.5 \\
2 & 0.02 & 7.41 & 9.11e-09 & 5.57 & 0.0188 &   6.1& 4.2e+04\\
3 & 0.03 & 8.82 & 1.68e-08 & 6.81 & 0.0371 &  17.8& 1.57e+03\\
4 & 0.05 & 8.92 & 2.99e-08 &  7.4 & 0.0638 &  34.3&   398\\
5 & 0.09 & 8.34 & 5.03e-08 & 7.92 & 0.102 &  50.8&   150\\
6 & 0.16 & 7.62 & 7.95e-08 & 8.57 & 0.156 &  63.6&  69.5\\
7 & 0.26 & 6.99 & 1.22e-07 & 9.37 & 0.236 &  72.6&    37\\
8 & 0.41 & 6.56 & 1.79e-07 & 11.1 & 0.365 &  78.6&  21.7\\
9 & 0.63 & 6.46 & 2.62e-07 & 14.8 & 0.601 &  82.9&  13.6\\
10 & 0.95 & 6.81 & 3.74e-07 & 25.1 & 1.33 &  85.9&  8.93\\
11 & 1.42 & 11.06 & 5.57e-07 &  153 & 10.5 &  88.3&   6.1\\
12 & *1.95 & *76.20 & *7.90e-07  & *2.01e+08& * 790&*  90.2&*   4.3 \\
13 & *2.77 & *53.12 & *1.22e-06  & *1.3e+08& * 617&*  91.9&*   3.0 \\
14 & *4.30 & *33.88 & *2.44e-06  & *6.49e+07& * 482&*  93.4&*   2.1 \\
15 & *11.21 & *12.84 & *3.27e-05  & *4.52e+06& *2.48e+03&*  93.1&*   1.3 \\
\hline
  \end{tabular}
\end{table*}

\begin{figure*}[hptb]
  \centering
  \includegraphics[width=12cm]{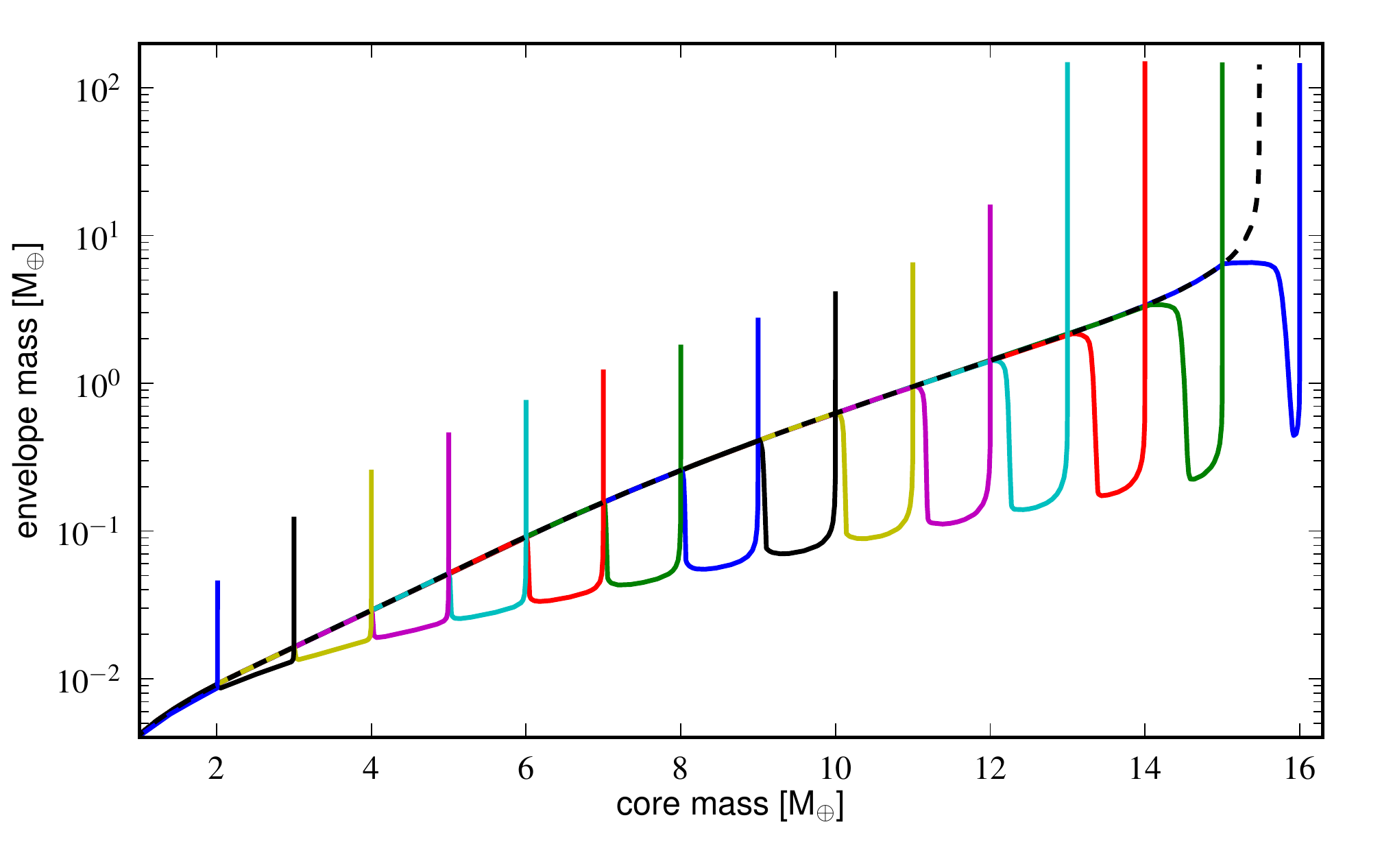} 
  \includegraphics[width=12cm]{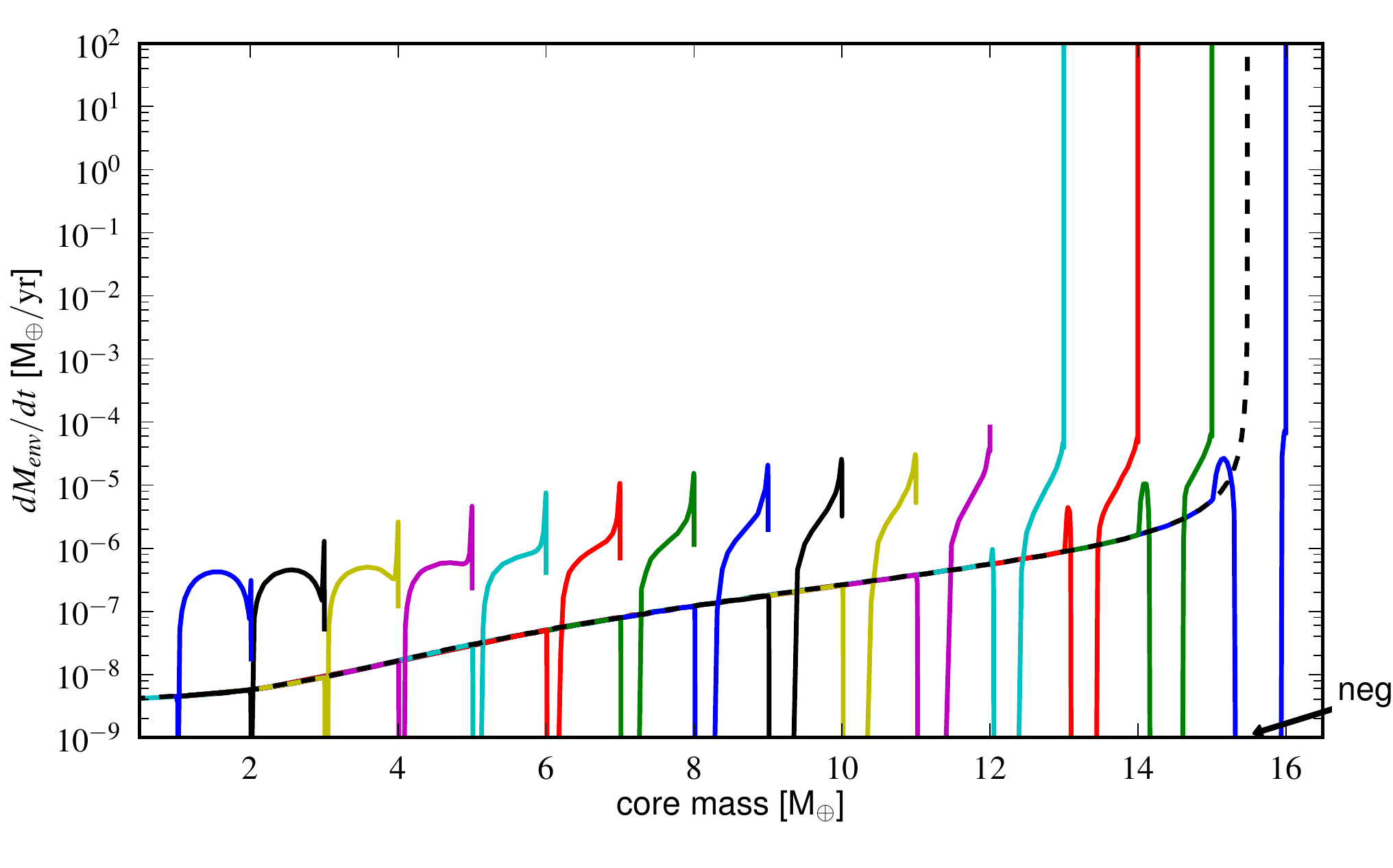} 
  \includegraphics[width=12cm]{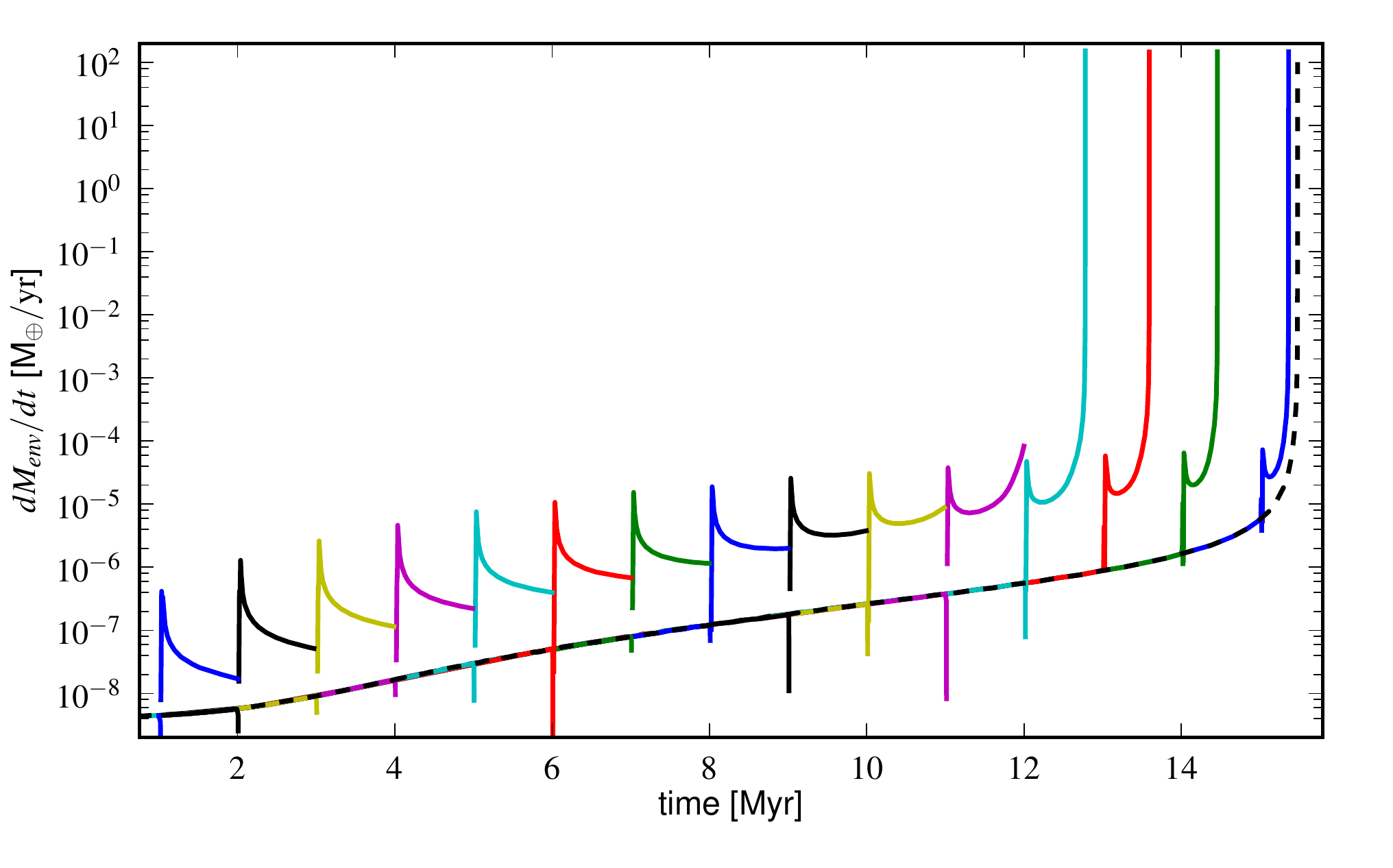} 
  \caption{Envelope mass and gas accretion rate following impacts of 1
    \mearth. The calculations with core 
    masses 12-15 end prematurely because gas accretion becomes
    supersonic. The dashed line represents the NC in all panels. The
    lower panels show the gas accretion rate in log-scale. Note that
    all values below $10^{-9}$ actually become negative. The lowest
    panel shows the envelope accretion rate as a function of time --
    the constant core accretion rate was used to convert the core mass
  to a proxy time. Each color represents a different planet having an
  impact that has evolved to this point by constant core growth.The
  very high accretion rate directly after the impact is similar to the
shut-off case. In case of shut-off the accretion rate rises
immediately but in the impact case, the impact and ejection causes a
delay (not visible on this scale). Afterwards the evolution is practically identical. For large
impacts, the henceforth larger core mass leads to a significantly
higher accretion rate when compared to the shut-off case.}
  \label{fig:i1}
\end{figure*}

For the most energetic impacts, Fig.~\ref{fig:i1} shows the envelope
mass and its accretion rate using different colors for all different
targets. For comparison, the NC is displayed as a dashed line. The
envelope mass as a function of core mass is represented in the upper
panel. We note that in all cases, the final envelope mass is
significantly larger than in the NC. The lower panels show the gas
accretion rate. Obviously, this assumed nominal rate has no relation
with physical reality, it only serves the purpose to compare  cores
after they have accreted the same amount of mass. In this figure one
can see that, after an initial phase of mass loss,  the gas accretion
rates stay always larger than the NC values in all cases. For core
masses between 12 to 15 \mearth, the impacts can even trigger rapid
gas accretion while the corresponding NC core has not yet reached its
critical mass. 

In the introduction we have said that rapid gas accretion after the
impact can be expected by analogy of the well studied shut-off
scenario. Our calculations confirm this. The clearly new and unexpeced
result of this study is the almost complete ejection of envelope and
the very fast reaccretion thereof after the impact. This
has strong implications for the composition of the envelope.

\section{Discussion}  
\label{sec:discussion}

We present a series of computations that aim at exploring the case in which episodic large bodies rather than a steady state of small ones provide the bulk of the core mass accretion. The resulting envelope structure, in particular the mass of the envelope, is compared between the two cases at a time when both cores have acrreted the same mass. We find that the large impacts or sudden energy input, while briefly reducing the envelope mass, allow for a larger gas accretion rate leading for all sizable impactors to a significantly more massive envelope. The resulting mass of the envelope normalized by the one obtained in the NC is listed in column 3 of Tables~\ref{tab:i002}-\ref{tab:i1}. Furthermore, the envelope is not only more massive but the gas accretion rate is also larger (column 5 of  the tables) indicating that the difference will keep growing. The exceptions are very small impacts on relatively large cores (0.1 \mearth on 13-15 \mearth cores and 0.02 \mearth onto 10-15 \mearth cores). In these cases, the impacts have little to no effect.  

\begin{figure}[btph]
  \centering
  \includegraphics[width=\imwidth]{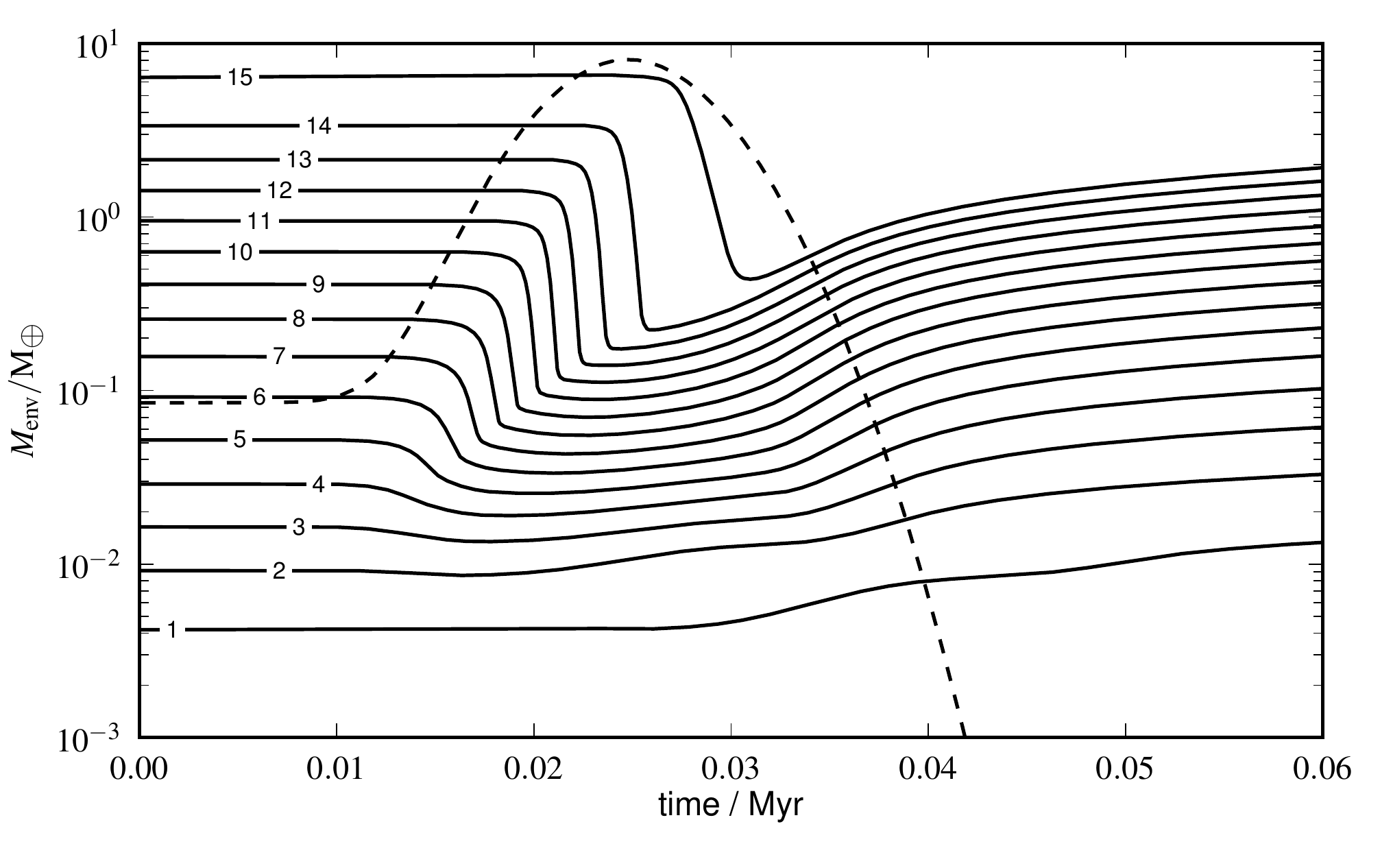}
  \caption{Envelope mass around the time of the impact  for a 1 \mearth
    impact.  The numbers give the target core mass in \mearth and the
    dashed line shows the core luminosity in arbitray units ($10^{21}$ W) to show
    when the "impact" takes place.}
  \label{fig:1e00env}
\end{figure}

To understand the reason for this, we focus  on the most dramatic 1 \mearth impact case. Figure~\ref{fig:1e00env} shows the envelope mass for all  different targets over 6 impact timescales. But now, time is plotted on the abscissa and we restrict the plot to 6 times the impact timescale. Each solid line stands for one target, the lines are labelled with the core masses. The dashed line shows the core luminosity in arbitrary units, to show the duration of the impact effects.
For small envelopes very little happens. Only after the impact, when the core luminosity becomes tiny, does the envelope mass increase slightly. In this case the envelope is so thin that the energy deposited can be transported very efficiently, therefore having no effect on the structure. As the envelope becomes more massive, the energy input starts to eject part of the envelope -- a significant fraction for envelopes larger than $3\cdot 10^{-3}\,$\mearth or cores larger than 3-4 \mearth. This effect becomes stronger as the envelope considered is more massive. One can also see, that a more massive envelope is ejected later in time. The reason for this is that it takes more time to inject enough energy to remove the envelope. In this case, the impact energy is always larger than the binding energy of the envelope. Therefore we eject a larger envelope fraction for larger envelopes because the energy transport becomes less efficient -- the energy is trapped in the envelope and can be used to remove it.

\begin{figure}[btph]
  \centering
  \includegraphics[width=\imwidth]{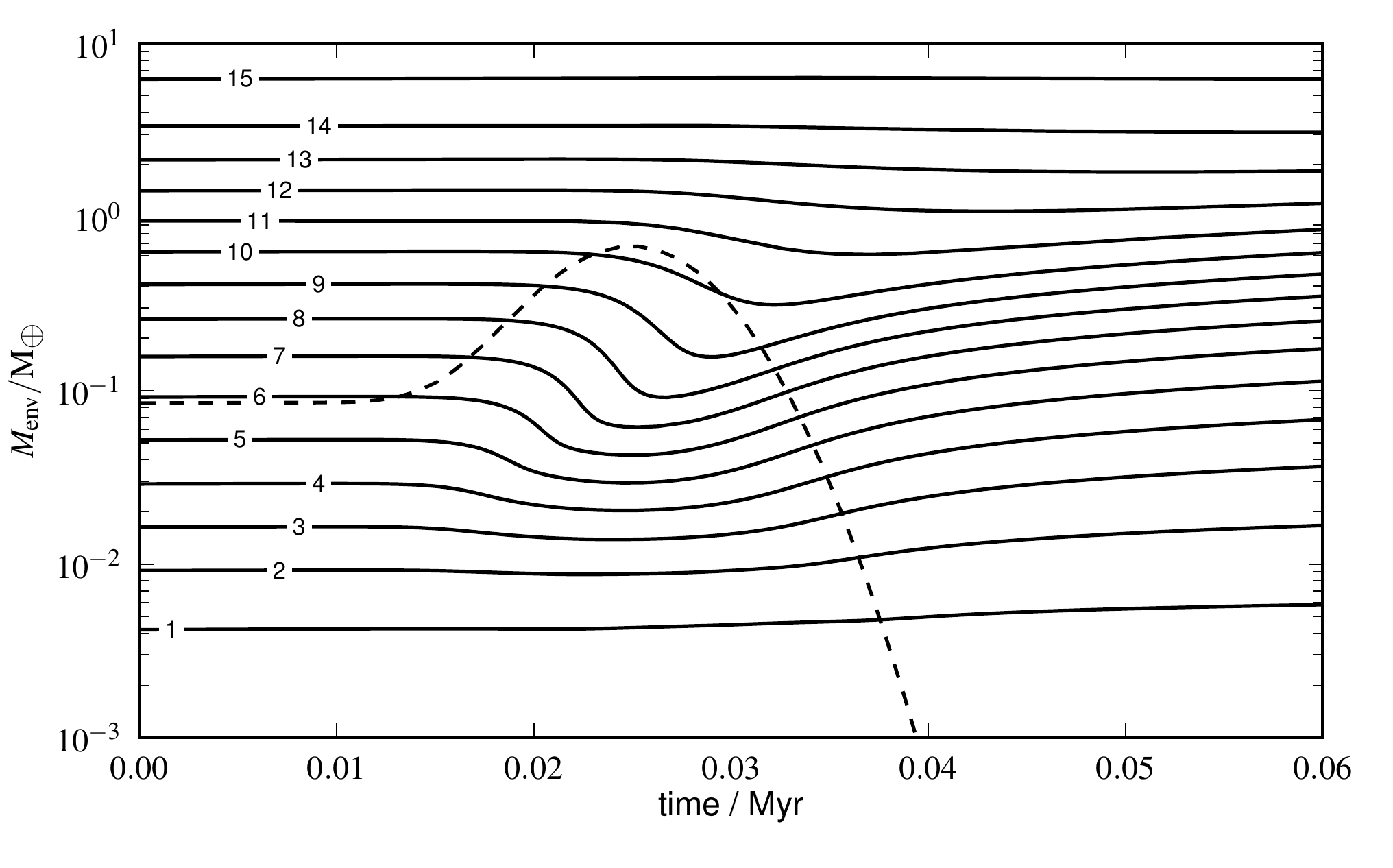}
  \caption{Same as Fig.~\ref{fig:1e00env} but for ten times smaller
    impacts (0.1 \mearth)}
  \label{fig:1e-1env}
\end{figure}

This is not always the case for smaller impacts, however. Consider the 0.1 \mearth impact case, Fig.~\ref{fig:1e-1env}.

\begin{figure}[tbph]
  \centering
  \includegraphics[width=\imwidth]{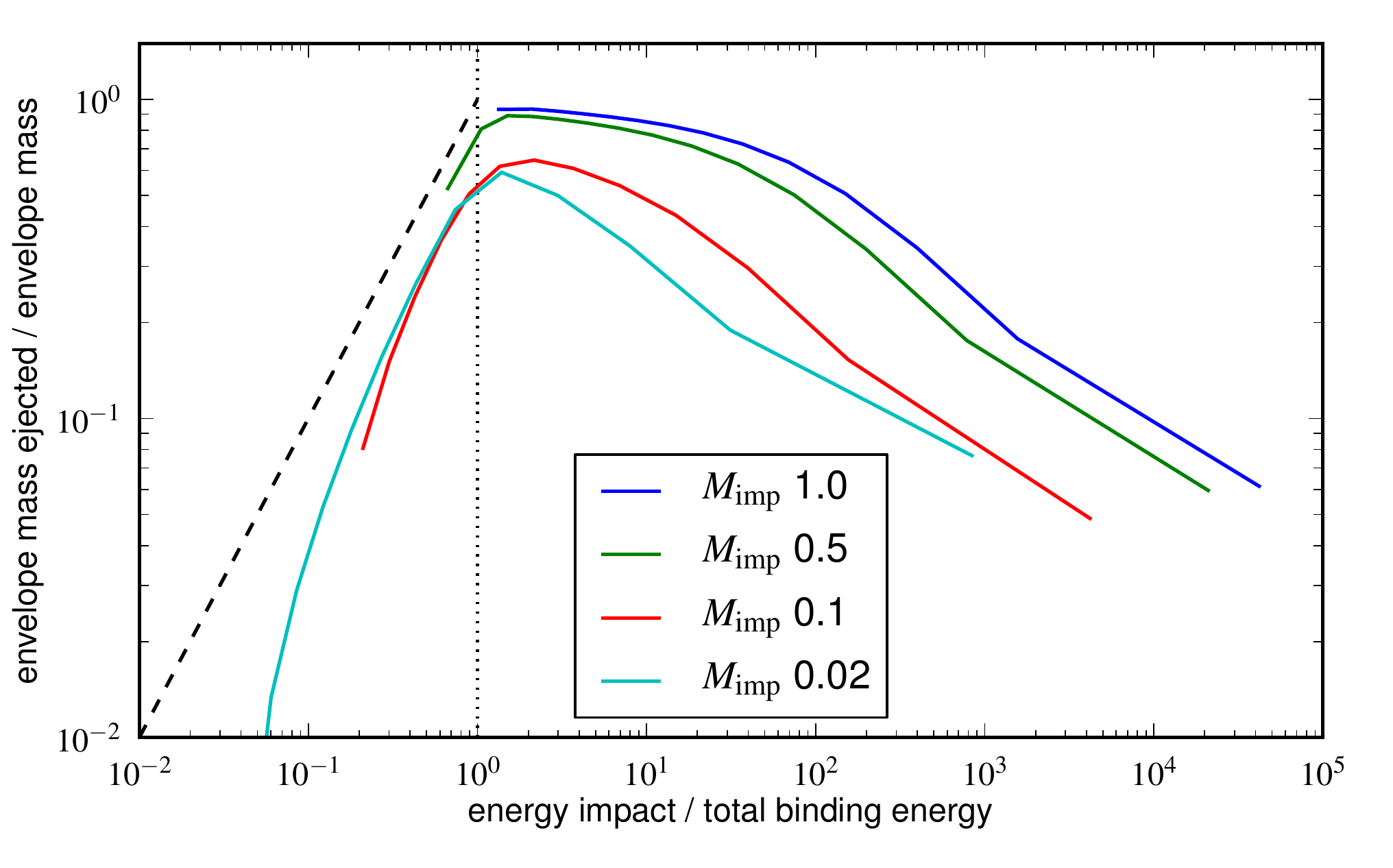}
  \caption{Fraction of envelope mass ejeced by the impact as a
    function of available energy. The abscissa shows the ratio of
    impact energy to total binding energy of the envelope. Left of the
  vertical dashed line the impact energy is insufficient to remove the
entire envelope.  The other dashed line shows
the ejected mass ratio for 100\% efficient processes. Note that large target cores are to the left, small
targets to the right.}
  \label{fig:env_ej_eimp}
\end{figure}

Starting with small envelopes, the behavior is similar. Only when the
envelope reaches a certain size, the envelope is partially
ejected. More so for more massive envelopes. Nevertheless, the
behaviour changes for very massive envelopes: less envelope is ejected
and eventually the impact has little effect. The reason is simple:
starting at cores of 10 \mearth and larger, the impact energy is
smaller than the total binding energy of the envelope (see
Table~\ref{tab:i01}). Therefore, while the process is efficient in
terms of using the available energy, there is not enough energy
available and only a fraction of the envelope can be removed from the
gravitational potential of the core and envelope.
The amount of available energy for all calculations is shown in
Fig.~\ref{fig:env_ej_eimp}. 
The largest impactor provides enough energy for all target
configurations, whereas the smaller impactors have insufficient energy
to eject the most massive envelopes completely. The dashed line shows
the enjected mass ratio for 100\% efficient processes. If the energy
of the impact is orders of magnitude larger than the binding energy,
this implies small targets with tiny envelopes. The impact energy can
be transported effecively and the envelope ejection becomes
small. That is why the curves go down again. In
this regime the impact shockwave might be efficient in removing the
entire envelope and a hydrodynamic treatment might be necessary (see
section~\ref{sec:impact-treatment-1}). For giant planet formation, this is
not a very interesting regime, however.

\begin{figure}[tbph]
  \centering
  \includegraphics[width=\imwidth]{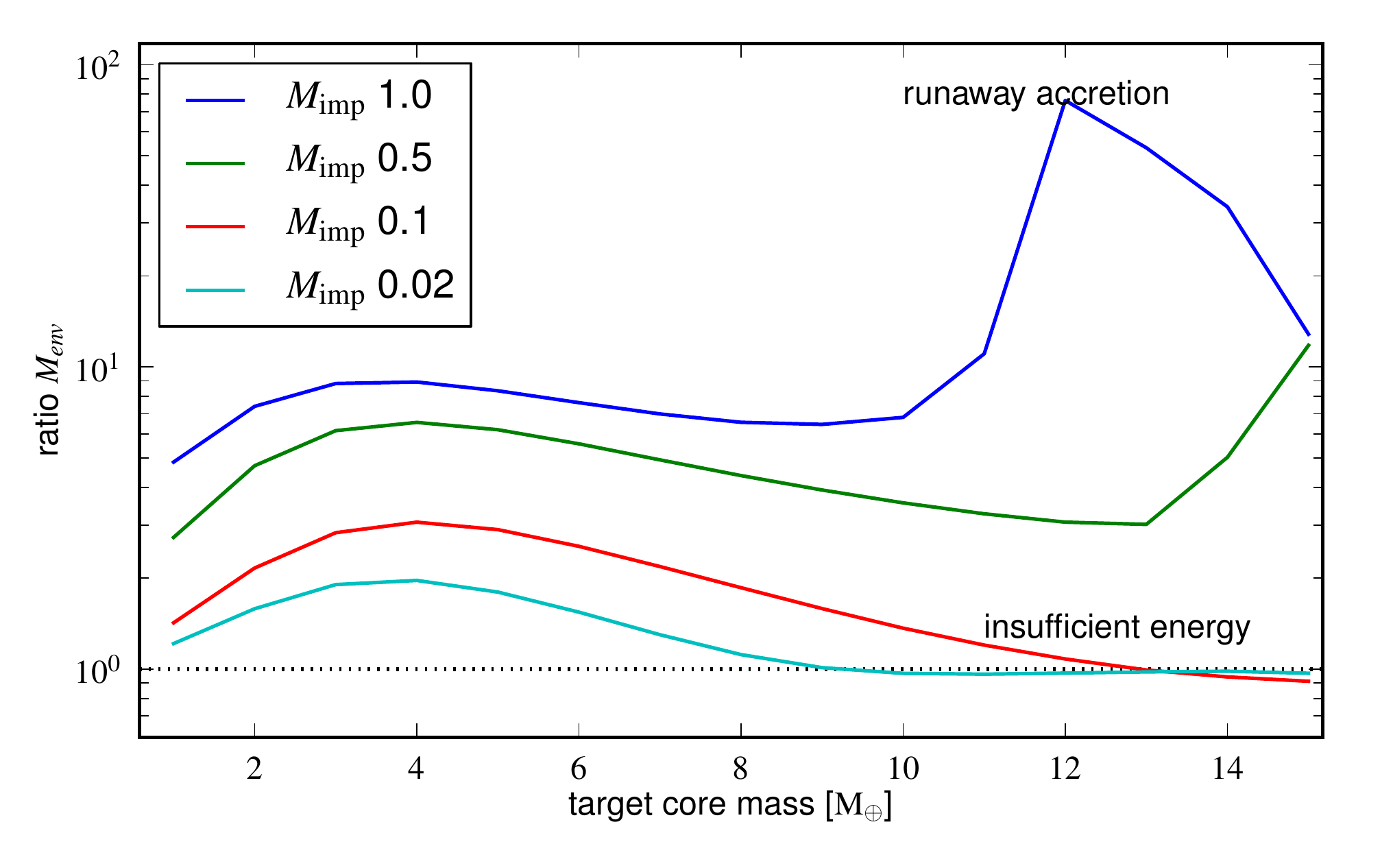} 
  \includegraphics[width=\imwidth]{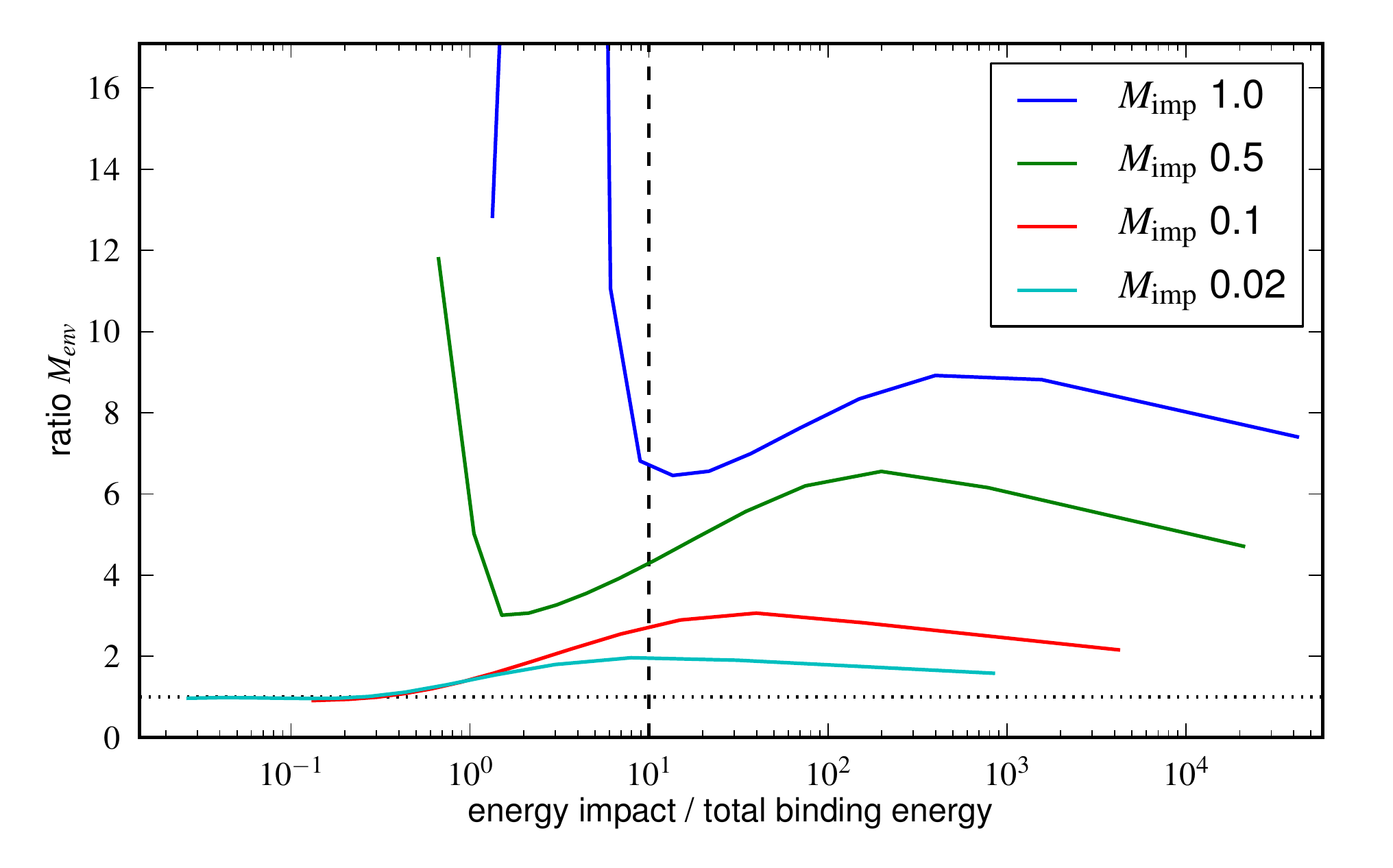} 
  \caption{Ratio of post-impact envelope mass to
    the NC envelope mass after the impact at \tcomp (same core mass). All
    computed impactors and target sizes are shown. top:
    abscissa shows the target core mass. Both large impactors provide
    sufficient energy to overcome the envelope binding energy for all
    targets. For the larger cores, runaway gas accretion is
    triggered. The smaller impacts do not have enough energy to
    significantly affect the large targets. bottom: abscissa shows
    impact energy as fraction of the total binding energy. Left of the
  vertical dashed line, the impact shockwave is not capable of
  ejecting the envelope directly (assuming that 10\% of the impact energy
 directly heats the envelope).}
  \label{fig:ratio_Menv}
\end{figure}

We have already said that the final envelope mass at \tcomp is larger
in the EC. To study this in more detail, we consider
Fig.~\ref{fig:ratio_Menv}. It shows the ratio of the envelope mass in
the EC to the NC (ratio $M_{env}$). The comparison time is chosen so that EC and NC have
equal core mass.
The top panel shows the envelope ratio vs. target core mass. For small
cores, this ratio rises with increasing core mass. For large cores,
runaway gas accretion can be triggerd if enough energy is available to
remove a large envelope fraction. If the energy is insufficient, the
envelope ratio decreases again. But only in the case of the smallest
impactor does it go below 1.

\begin{figure}[tbp]
  \centering
  \includegraphics[width=\imwidth]{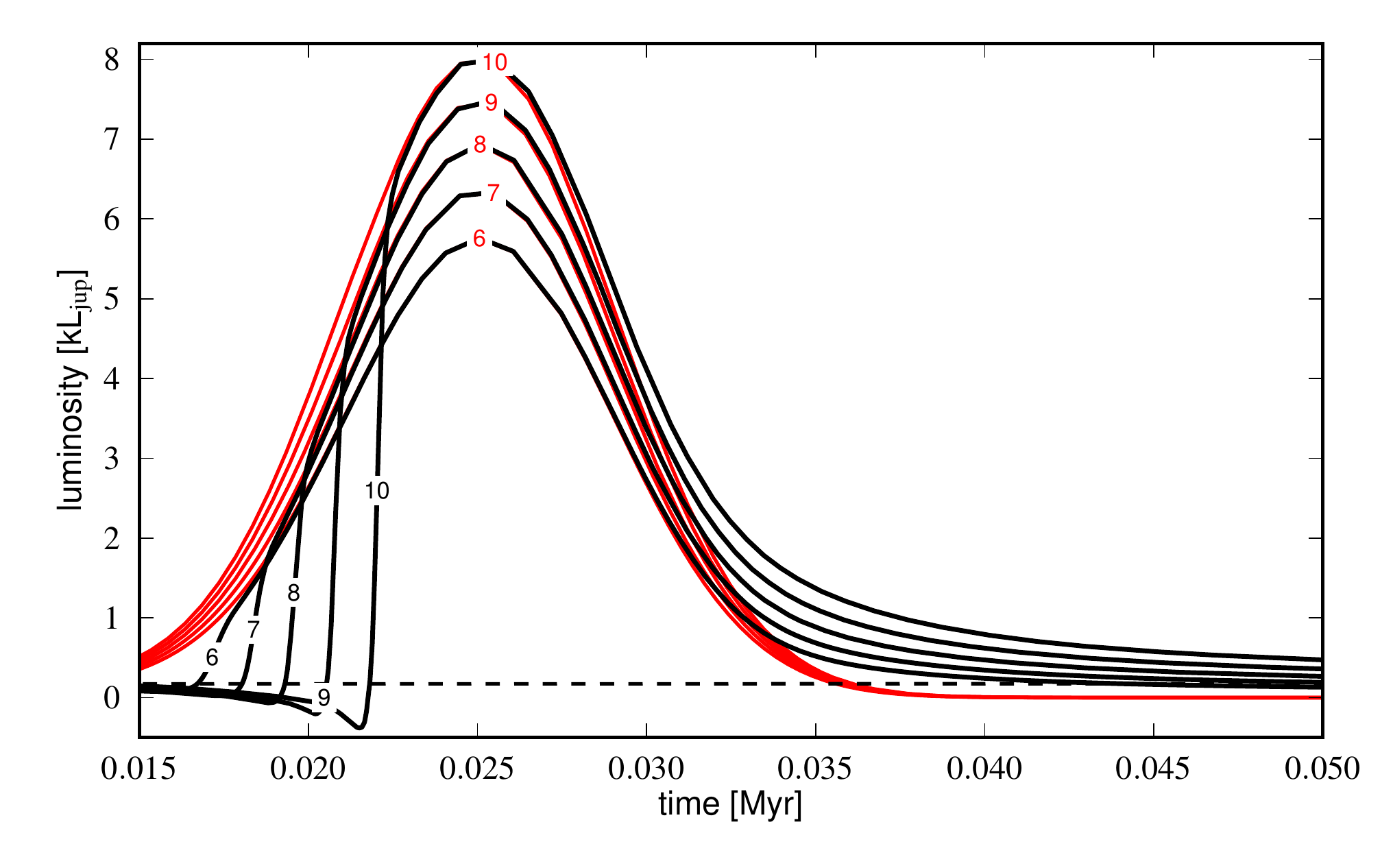} 
  \includegraphics[width=\imwidth]{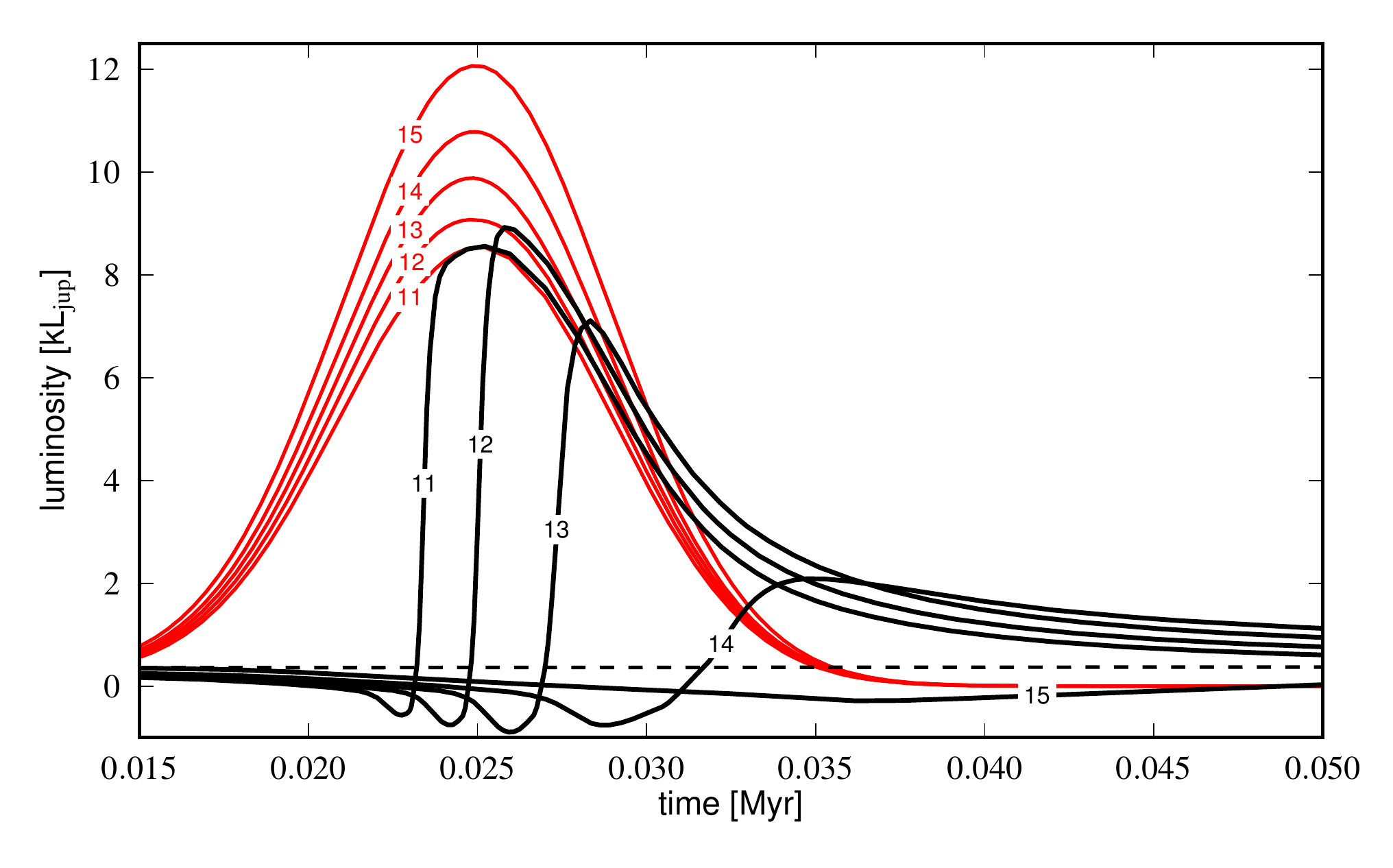} 
  \caption{Luminosity during impacts of 1 \mearth for 
    target core sizes from 6-10 \mearth (top) and 11-15
    \mearth (bottom). The thick black line shows the total luminosity
    and the thin red lines show the core luminosities. Small labels
    indicate the target core mass in \mearth. For
    reference, the dashed line shows the nominal case of gradual core
    growth for the most massive target.}
  \label{fig:luminosityexample}
\end{figure}

After discussing the energetics of the envelope removal, we consider
the acceleration of gas accretion. Only if the time for re-accretion of the
ejected gas is small compared to the periods
in-between impacts (the shut-off-like phase) we have a resulting net
speed-up. Why is the ejected gas replaced so fast? 
 It turns out that the initial envelope loss is a key
factor. The gas accretion is governed by the
rate by which the energy can be radiated away. For sub-critical
cores, the luminosity or planetesimal accretion rate sets the internal
structure of the envelope and in the static limit directly the total
envelope mass. When an
impact occurs, the energy is used to eject a large
fraction of the envelope and the remaining energy is radiated away
very fast due 
to the high luminosity. The high luminosity is possible because of the
tiny envelope. Afterwards, the core luminosity is low and
gas accretion can be very fast. All the accretion energy that is
evenly spread out in the NC has been
radiated away or used to eject the envelope.
Figure~\ref{fig:luminosityexample} shows the time evolution of the
luminosity for 0.5 \mearth impacts onto intermediate cores. It shows
both the initial decrease in total luminosity $L$, and the very large
value of $L$ shortly afterwards. After the envelope has expanded,
and after the impact energy has been transported to the surface, the
luminosity is mainly produced by contraction of the
envelope. Therefore the luminosity quickly shrinks to and below the
nominal value representing constant core accretion. This is not
visible in Fig.~\ref{fig:luminosityexample} because the figure only
shows the first 0.1 Ma after the impact to resolve the luminosity
curve. Afterwards,
there is still a long time (in this case 0.5 Myr) to accrete envelope
gas with a very small core luminosity.

For more extended envelopes, another effect comes into play: the
binding energy of the envelope. When the mass of the
envelope becomes important, injecting energy at the center of the
planet will not lead to an increase of temperature. On the contrary, it 
 will actually reduce the temperature after a very brief temperature
 increase. This is caused by the  negative gravothermal specific heat
 which is well-known for stars 
where is is responsible for stable hydrogen burning. When nuclear energy
production suddenly increases, the temperature will decrease and
reduce the nuclear energy production. Following \citet{kippenhahn}
(section 25.3.4) we define the
gravothermal specific heat as:
\[
c^* = c_P \left(1-\nabla_{s}\frac{4\delta}{4\alpha-3}\right)
\]
where $c_P$ is the specific heat at constant pressure and $\alpha$ and
$\delta$ are the generally used logarithmic derivatives in the equation
of state of density
vs. pressure or temperature, respectively. If $c^*$ is negative,
adding energy $dq$ to the gas will reduce its temperature:
 more
energy is needed for the expansion of the envelope than what is
originally the cause for the expansion. Applying this definition to
our envelope structures prior to impact shows that $c^*$ is indeed
negative for the central part of the envelopes. 
The effect is even stronger for gas giant planets
than for stars (of comparable size), because the gravitational field of the core enhances
the effect that is only caused by the self-gravity of the gas for
stars. 
 In our calculations, this effect can be seen by
the decrease in luminosity in the initial impact phases. For massive
envelopes, this effect is so strong that 
a temperature inversion happens in the outer envelope and the total
luminosity becomes negative for a short time. During this phase, the
planet cannot radiate its internal energy away. But this phase lasts
only a very short time (e.g., 1200 yr in the $M_c=8$ \mearth, 1 \mearth
impact case). After this phase, the planet is relatively cold and can
accrete gas faster than before the impact. 

 This is the reason why the lost
 envelope mass can be replaced with new gas so fast that the envelope
 ejection 
 has no important consequences for the evolution of the planet, see
 section~\ref{sec:comp-with-stopp}.

\subsection{Consequences of the impact shock}
\label{sec:impact-treatment-1}
A realistic impact for relatively small envelopes that we are
considering can be divided in the following stages: 1) the
impactor traverses the envelope and deposits some energy directly in
all layers. 2) The impactor hits the core and deposits a fraction of
its energy directly in the lower layers of the envelope. This causes
a blast wave to travel through the envelope. 3) The energy deposited
in or around the core is transported to the envelope and the envelope
reacts to this. 4) After the impact energy has been processed, the
envelope reacts to the decreased energy flux and typically contracts.

In our computation, we neglect stage 1. 
Stages 2 and 3 are modelled together
in a simplified fashion by assuming that the energy is released on a
certain timescale.
Test calculations show that the the reaction of the
envelope does not depend strongly on the exact value of this timescale
as long as two things are true: it is much longer than a dynamic
timescale so that the envelope has enough time to stay in hydrostatic
equilibrium. Secondly, the timescale must be much shorter than the
cooling timescale. This means that there is no time for the envelope
to radiate the energy during the "impact". Stage 4, the gas
accretion after the "impact" is modelled consistently. 

This treatment neglects the fact that, for large impacts, stage 2 can
dynamically strip the envelope. This has been shown by
\citet[][h.f. KOR90]{1990Icar...84..528K}. 
We have also begun to simulate this impact with 3-dimensional SPH
simulations using dunite for the core and ideal gas for the
envelope. We were using a SPH formulation suitable for handling high
density contrasts. 
Our calculations show a gradual change from no
ejection to full ejection of envelope and a strong dependence on
impact paramenter:  Large impact angles are more likely and tend to
eject less  
envelope. We will study this in detail in a forthcoming article
(A. Reufer et al., in prep.). Nevertheless, it appears clear that for giant
impacts significant fractions of the envelope can be ejected,
especially on smaller targets.
 However, in such a case, impact
energies are large and subsequently stage 3 will continue to eject the
(remaining) envelope and prevent envelope accretion due to huge
luminosity until 
stage 3 has ended and core luminosity drops back to 'normal'
levels. Afterwards the behaviour will be similar independent of the
amount of gas ejected already during stage 2. Therefore we argue
that the mid- and long-term outcome will be independent of the exact
nature of envelope ejection. Note that during stage 3, near the peak
core luminosity, the envelope is again in a static state with
$l=const$ throughout the envelope in most cases (see
Fig.~\ref{fig:luminosityexample}). This means that it has 'forgotten'
its history and previous envelope ejection -- by blast wave or
otherwise -- is irrelevant.

\subsection{Comparison with a stop of core accretion}
\label{sec:comp-with-stopp}

\begin{figure}[tbhp]
  \centering
  \includegraphics[width=\imwidth]{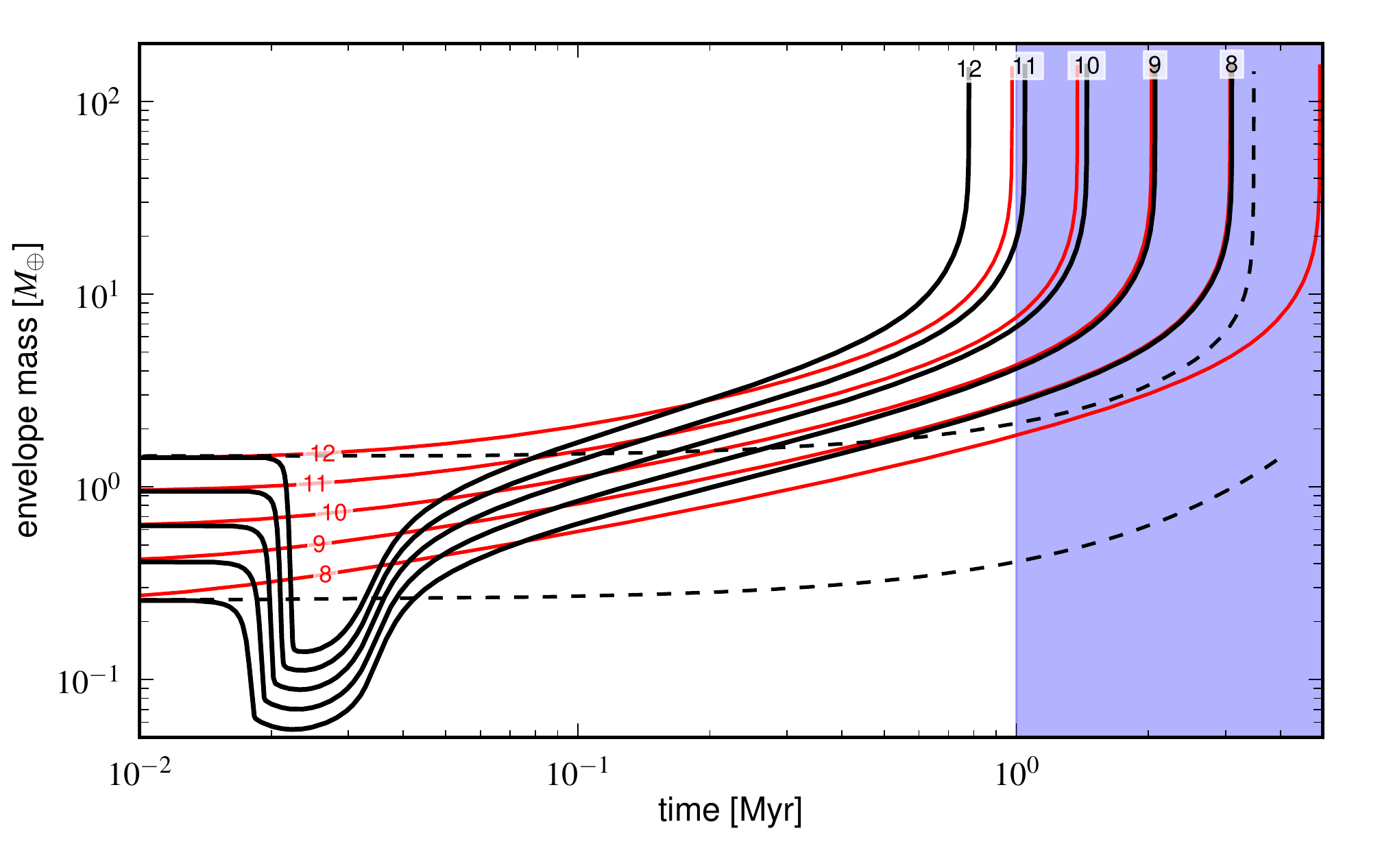}

  \caption{Gas accretion compared to shutdown scenario. Black lines
    show the envelope mass after 1 \mearth impacts on cores of 8 - 12
    \mearth. Red lines show the evolution when
    the core accretion rate is simply shut off with the same timescale
    and to the same background rate instead of impacting 1
    \mearth. Dashed lines show the envelope mass of the NC for the
    largest and smallest target. At 1 Myr the NC and the impact case
    have the same core size. Cases 8-11 have been extended beyond the
    1 Myr \tcomp that was usually adopted to compare to the NC.}
  \label{fig:compshut}
\end{figure}

 \cite{2000ApJ...537.1013I} and \citet{Hubickyj2005415} have already studied the effect
of shutting off core accretion on the gas accretion rate. Here we compare
the impact to core luminosity shut-off.
Figure~\ref{fig:compshut}
shows core masses in the range from 8 to 12 \mearth and the evolution
of the envelope mass in 1) the impact case and 2) when turning off
core accretion. We use the same timescale of $10^4$ years to reduce
the core accretion rate to the background level. To allow a direct
comparison, we have extended the old calculations to later times to
see when the planets start rapid gas accretion. The solid lines show
the impact scenario, while the red lines represent shutdown of core
accretion. 

It is evident, that the shutdown leads to a quicker envelope build-up
initially. The impact cases first expand and loose a significant mass
fraction. After the envelope loss, however, the impact cases quickly gain
envelope mass, much faster than the shutdown case. At all times after the initial ejection, the
envelope accretion rate stays larger for the impact case. For the
comparison it is important to note that the impact case has a heavier
core by 1 \mearth compared to the shutdown case. When comparing with
shutdown cases 1 Myr later (having the same core size), the envelope
accretion rates become similar only after the planet has "forgotten" the
impact. This occurs after a few Kelvin-Helmholtz timescales\footnote{Note that the
  Kelvin-Helmholtz timescales shortly after the impact are quite
  short, e.g. the $M_c=10\, \mearth$ case, when the accretion rate has
  reached the background level, has a
  Kelvin-Helmholtz timescale of 0.05 Myr. The smallest value during the
  impact is as low as 180 yrs. The  $M_c=10\, \mearth$ shutdown case has a
  Kelvin-Helmholtz timescale of 0.17 Myr directly after shutdown.}. In
the end the planets reach an envelope mass of the same size as the
respective larger core in the shutdown case.

In other words, the impact scenario allows gas accretion at a slightly
enhanced rate compared to the shutdown case while concurrently growing
the core. When comparing cores of equal size, the impact case is
almost as fast as the shutdown case. Only for very large cores, when
the time-to-runaway in the shutdown case is of the order of 1 Myr or less
(i.e. the time it takes to gradually increase the core by the size of
the impactor is longer than the time to reach rapid gas accretion), is
the impact case slower than the shutdown case. 
For long stretches between impacts, the absence of core
accretion is the important effect but the reconfiguration of the
envelope during the impact plays an important role shortly after the impact.
Hence, when the times between impacts are relatively short, the
impact case is still faster than the shutdown case (for equal initial
core mass). This is only
changed when the time between impacts becomes so short, that the
initial mass loss cannot be compensated anymore.

\section{Conclusion}  
\label{sec:conclusion}

We have analyzed the effect of relatively massive impacts onto the
cores of giant planets in the growth phase. Previous studies
\citep{2000ApJ...537.1013I,Hubickyj2005415} have shown how the
envelope reacts to a shut-off of core luminosity with rapid envelope
accretion. Therefore, intermittent giant impacts with long periods of
very low core accretion in-between are expected speed up gas
accretion. The aim of this study was to study the reaction of the
envelope to the giant impact and the subsequent period of low core
luminosity in detail. Wether or not a net acceleration of gas accretion takes
place and how strong this effect is, depends on the timescale of the
reaction to the impact in relation to the time in-between impacts.

There are two major effects: 1) Due to a very
large core luminosity after the impact, a large fraction of the
envelope is ejected. The remaining tiny envelope allows a very fast
energy transport and in consequence the huge impact energy is used up
or radiated away very fast. The huge luminosity in combination with a
small envelope leads to a small Kelvin-Helmholtz time. Therefore the
envelope 'forgets' that the impact has taken place in very short
time. Afterwards, very little solid accretion is
necessary for a given net core growth speed and the subsequent
evolution up to the next impact can be understood by the well-studied
sudden shut-off of core luminosity. However, due to the episodic impacts, further core growth
is still possible. 
2) This effect is enhanced by a second: The very
large luminosity during the impact reconfigures the envelope structure
so that it is in hydrostatic equilibrium for very high energy
fluxes. Shutting off core accretion after this reconfiguration
triggers higher gas accretion rates than the shut-off without prior
impact. This can also be understood in terms of the negative
gravothermal specific heat of self-gravitating non-degenerate
gases. In this state the impacts actually lower the central
temperature which explains why subsequent gas accretion can be faster
after the impact. This second effect helps reduce the time it takes to
accrete once more the envelope gas that has been ejected by the
impact. Once the planet has again the same envelope mass as before the
impact but a much lower core luminosity, the evolution follows the
shutdown scenario.

Together, the
alternation between very high and low energy input allows more gas to
be accreted even though the impact initially ejects some (or all) of
the envelope gas. The subsequent high gas accretion rate quickly
reforms the original envelope and continues to accrete faster than the
gradually growing case. In this way, even a rapid succession
of impacts can lead to a faster envelope growth.

A further interesting consequence of the envelope loss caused by the
impact relates to the dust opacity: every time the envelope is
ejected, new fresh gas is accreted from the nebula, therefore
resetting any former modifications to the opacity caused by e.g. dust growth and settling.

In summary, we find that episodic large impacts significantly speed up
gas accretion as was expected from shut-off calculations. The new
result of this study is the fact that 1) almost the entire envelope of
the planet is ejected as a consequence of the impact energy and 2) it
takes only a very short time to accrete the lost gas after the
ejection. In fact, this is so fast that the ejection has practically
no effect on the long-term evolution - envelope masses are 
equivalent with the shutdown-case using the increased post-impact core mass.

We can therefore conclude the following for the formation of giant
planets: If planetesimals are accreted in a regime where mass ratios
are large, i.e. most mass is delivered by massive impacts, this will
accelerate envelope build-up. Furthermore, the ratio of envelope to
core mass will be significantly enhanced and smaller cores can begin
rapid gas accretion while the core is growing by large impacts. 
The speedup with consecutive impacts, can be understood in principle from these calculations. We will study a full evolution
calculation based on episodic core growth in a future publication.


\section*{Acknowledgements}
CB wishes to express his thanks to a number of people who have
contributed to this work, especially to realize the new code: S. Krause wrote the first version of the
implicit code for ideal gas. G. Wuchterl assisted in the code design
and in particular was responsible for consistent spline interpolation
of SCVH equations and opacity tables to include realistic materials. Also,
his help was greatly appreciated for the finite-volume discretization of the physical equations.
A. Reufer did the SPH simulations of the impact and administrated the computer cluster where all
these calculations have been performed. Special thanks go to the
annonymous referee who pressed to clarify the differences to the
shut-off scenario which made the results much clearer.
Part of this work has been supported by the Swiss National Science Foundation.

\bibpunct{(}{)}{;}{a}{}{,} 
\bibliographystyle{aa}
\bibliography{bib/Literatur}







\appendix

\section{List of symbols}
\label{sec:list-symbols}

\begin{table}[htb]
  \centering
  \caption{List of symbols}
  \label{tab:symbols}
  \begin{tabular}{@{$}l@{$\hspace{1ex}}ll}
    \hline
\textrm{name} & unit & description \\\hline
a & m & semi major axis\\
a & $\rm J\, m^{-3}\, K^{-4}$ & radiation  constant\\
c_P & $\rm J\,kg^{-1}\,K^{-1}$ & specific heat at constant pressure\\
\epsilon_{ac}& $\rm J\, kg^{-1}\, s^{-1}$ & energy in-flux from
planetesimal accretion per unit mass\\
F & $\rm W \, m^{-2}$ & energy flux\\
\Phi & $\rm J\, kg^{-1}$ & gravitational potential\\
\delta & 1 & equation of state: $\delta\equiv - \partial \ln \rho / \partial \ln T |_P$ \\
G & $\rm N \, m^2 \, kg^{-2}$ & gravitational constant\\
\kappa & $\rm m\,kg^{-1}$ & Rosseland-mean opacity\\
l & $\rm W\, m^{-2}$& luminosity\\
m & kg & envelope mass\\
M_c & kg & core mass\\
M_{\ast} & kg & host star mass\\
n & 1 & point concentration\\
\nabla & 1 & log. temperature gradient $d \ln T/d\ln P$\\
\nabla_s & 1 & isentropic temperature gradient $d \ln T/d\ln P|_s$\\
P & Pa & pressure\\
q & $\rm J\, kg^{-1}$ & heat per unit mass\\
T & K & temperature\\
r & m  & radius \\
\rho & $\rm kg \, m^{-3}$ & density\\
\sigma & $\rm W \, m^2 \, K^{-4}$ & Stefan-Boltzmann constant\\
t& s & time\\
\tau_{EW} & s & impact timescale (equivalent width of Gaussian) \\
\theta&1&time centering parameter 0..1\\
u_{rel} & $\rm m/s$& relative velocity of matter with respect to grid:
$u_{rel} + u_{grid} = u$\\ 
V & $\rm m^3$ & volume\\
e &  $\rm J\, kg^{-1}$ & specific internal energy per unit mass \\\hline
  \end{tabular}
\end{table}

\begin{table}[htb]
  \centering
  \caption{Operators}
  \label{tab:operators}
  \begin{tabular}{lll}
\hline
  operator & description & definition \\\hline
$\delta$ & temporal delta  &$\delta x = x^{new} - x^{old}$\\
$\Delta$ & spacial delta on scalar & $\Delta S_j=S_{j+1}-S_j$\\
& spacial delta on vector & $\Delta v_j=v_{j}-v_{j-1}$\\
$\widehat{\hspace{1ex}}$ &time-centering&$\widehat{x}=\theta
x^{new}+(1-\theta) x^{old}$\\
$\widetilde{\hspace{1ex}}$& advection & donor cell: take upstream value
\\\hline
  \end{tabular}
\end{table}

\begin{table}[htb]
  \centering
  \caption{List of non SI units with adopted values}
  \label{tab:units}
  \begin{tabular}{lll}
\hline
  unit & description & value \\\hline
\mearth & mass of Earth  & 5.9742e24 kg\\
$L_{\mathrm{jup}}$ &Jupiter internal luminosity   & 3.35e17 W  \\
yr & julian year   & 31557600 s 
\\\hline
  \end{tabular}
\end{table}

\end{document}